\documentclass[11pt]{article}

\usepackage{amsmath}
\usepackage{amsfonts}
\usepackage{amssymb}
\usepackage{graphicx}
\usepackage{supertabular}
\usepackage{setspace}

\RequirePackage{cite}%
\renewcommand{\citeleft}{\bgroup\normalfont[}%
\renewcommand{\citeright}{]\egroup}%

\newcommand{\n}{\noindent}
\usepackage{mathptmx}

\newtheorem{theorem}{Theorem}
\newtheorem{corollary}{Corollary}

\begin{document}

\title{\leftline{\small Published in General Relativity and Gravitation (\texttt{www.springerlink.com})}
\leftline{\small M. Azreg-A\"{\i}nou, \textit{Gen. Relativ. Gravit.}, DOI 10.1007/s10714-009-0915-6}
\vskip0.5cm
Selection criteria for two-parameter solutions to scalar-tensor gravity}
\author{Mustapha Azreg-A\"{\i}nou
\\
Ba\c{s}kent University, Department of Mathematics, Ba\u{g}l\i ca Campus, Ankara, Turkey}
\date{}

\maketitle

\begin{abstract}
We make a systematic investigation of the generic properties of static, spherically symmetric, asymptotically flat solutions to the field equations describing gravity minimally coupled to a nonlinear self-gravitating real scalar field. Seven corollaries and a theorem on selection criteria for two- and one-parametric solutions are proven and conditions for obtaining particle-like solution, black holes or naked singularities are derived. A series of exact solutions in closed forms describing different black holes, naked singularities and particle-like solutions are provided.

\vspace{3mm}

\n {\footnotesize\textbf{keywords.} Einstein-scalar gravity; black holes; naked singularities; particle-like solutions; exact solutions}

\vspace{-3mm} \n \line(1,0){431} 
\end{abstract}

\section{Introduction}
Due to the accelerated expansion of the Universe, scalar field dark matter models emerged to account for the astronomic observations (see~\cite{rev} for a review). The models describe dark matter as a scalar. These models are special issues of more general theories known as scalar-tensor theories. In contrast with pure gravity, four-dimensional scalar-tensor gravity admits a variety of \textit{static}, spherically symmetric, neutral or charged configurations including Schwarzschild black holes (BH), phantom BHs, naked singularities (NS), particle-like (P-L) solutions and wormholes~\cite{B07,B06,N08,B01D,B01S,O95,C95}.

We consider the four-dimensional action of general relativity coupled to a self-gravitating real scalar field with positive kinetic energy (Einstein-scalar gravity)
\begin{equation}\label{S}
    S=\int \text{d}^4x\sqrt{-g}\,[R+\phi_{,\mu}\phi^{,\mu} - 2V(\phi)]\,,
\end{equation}
where $R$ is the scalar curvature and $V(\phi)$ is a potential for $\phi$. The field equations derived from~\eqref{S} are~\cite{B01D}
\begin{equation}\label{d11}
    \phi'\nabla^{\mu}\nabla_{\mu}\phi + V' = 0\,,\quad R^{\mu}{}_{\nu} - \frac{1}{2}\,\delta^{\mu}{}_{\nu}R + T^{\mu}{}_{\nu} = 0\,,
\end{equation}
(where $'\equiv \text{d}/\text{d}\rho$), with $T^{\mu}{}_{\nu}$ being the conserved energy-momentum tensor of $\phi$
\begin{equation}\label{em}
    T^{\mu}{}_{\nu} = \phi_{,\nu}\phi^{,\mu} - \frac{1}{2}\,(\phi_{,\sigma}\phi^{,\sigma})\delta^{\mu}{}_{\nu}
    + V(\phi)\delta^{\mu}{}_{\nu} \,.
\end{equation}

For static, spherically symmetric configurations, the metric can be put in the form
\begin{equation}\label{m}
    \text{d}s^2 = A(\rho)\,\text{d}t^2 - \frac{\text{d}\rho^2}{A(\rho)} - r^2(\rho)\,\text{d}\Omega^2\,.
\end{equation}
We fix the behavior of $r(\rho)$ at spatial infinity such that $r\simeq \rho$ as $\rho\to \infty$ (this is no loss of generality). With the ansatz~\eqref{m}, the first equation in~\eqref{d11} reduces to Eq~\eqref{f1} and the second group of equations in~\eqref{d11} lead to Eqs~\eqref{f2} to~\eqref{f4}~\cite{B01D}
\begin{eqnarray}
\label{f1}  & &r^2V' = \phi'(Ar^2\phi')'\,, \\
\label{f2}  & &V = -(A'r^2)'/(2r^2)\,, \\
\label{f3}  & &\phi'^{\,2} = -2r''/r\,, \\
\label{f4}  & &A(r^2)'' - r^2A'' = 2\,,
\end{eqnarray}
Only three equations of this set are independent; when the last three equations, \eqref{f2} to \eqref{f4}, are satisfied the first one is automatically satisfied.

\begin{figure*}
  \includegraphics[width=0.7\textwidth]{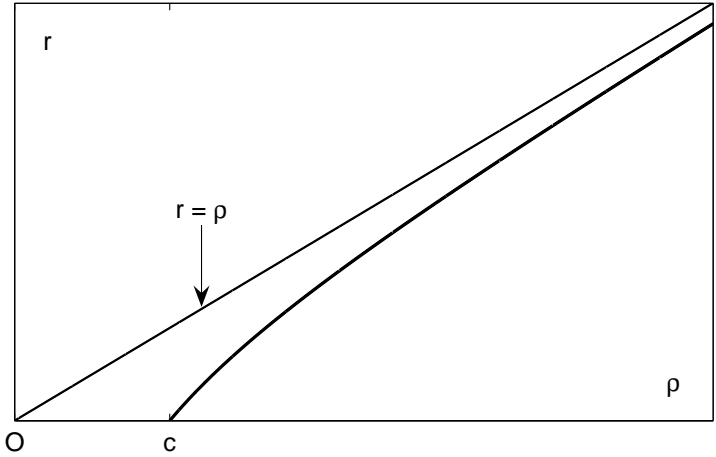}
  \caption{Typical behavior of $r(\rho)$. $c$ is a scaling constant and location of the center of configuration.}\label{Fig1}
\end{figure*}

Theorem 1 of Ref.~\cite{B01D} and its generalization to $(d>4)$-dimension~\cite{B01S} rules out wormhole, horn and flux tube type solutions to Eqs~\eqref{f1} to \eqref{f4}. A consequence of that, the range of $r$ is from $r=0$ at the center to $\infty$; correspondingly, the range of $\rho$ is from $\rho=c$ at the center of configuration ($r(c)=0$) to $\infty$. Since $\phi'^{\,2}\geq 0$ (real scalar field), \eqref{f3} leads to $r''\leq 0$. Hence, the function $r(\rho)$ is concave down on its range. Fig.~\ref{Fig1} shows a generic typical behavior of $r(\rho)$ where $c$ is the location of the center. In this paper $c$ is treated as a(n arbitrary) scaling constant (which can be set to 1) to which we compare the masses of the solutions and the whole theory is scale invariant\footnote{Setting $c=1$ is the same as the rescaling $\rho/c\rightarrow \rho$, $m/c\rightarrow m$, $A\rightarrow A$, $\phi\rightarrow \phi$, $c^2V\rightarrow V$.}. $c$ is also a measure of the strength of the field. We also make it such that the value of $\rho$ at $r=0$ is precisely $c$; this is no loss of generality as justified in Appendix A.

For the case of four-dimensional field equations~\eqref{f1} to~\eqref{f4}, Bronnikov~\cite{B01D} established some restrictive results, which have been latter generalized to $d$-dimensional field equations~\cite{B01S}. We are concerned with the following conclusions~\cite{B01D,B01S}.
\begin{itemize}
  \item All horizons are simple.
  \item All possible global structures for $\phi\neq \text{constant}$ and for $\phi= \text{constant}$ are the same;
  \item Possible non-singular solutions have either Minkowski/Ads or de Sitter structures. P-L asymptotically flat solutions with mass $m>0$ and $V$ having both signs are not excluded; however, if $V\geq 0$ no such P-L solutions ($m>0$) exist.
  \item For all functions $V(\rho)$, Eqs~\eqref{f1} to~\eqref{f4} do not admit static, spherically symmetric regular BH and wormhole solutions with any asymptotic.
  \item For $V\geq 0$, Eqs~\eqref{f1} to~\eqref{f4} do not admit any hairy BH solutions (no scalar hair theorems~\cite{beken}).
\end{itemize}

An asymptotically flat BH solution to Eqs~\eqref{f1} to \eqref{f4} is characterized by the presence of event horizons. Since all horizons are simple, there is a change in the sign of $A(\rho)$ at each event horizon. Thus for a BH, $A(\rho)$ has both signs in the domain $c\leq \rho <\infty$. For an asymptotically flat P-L or NS solution there are no event horizons, thus $A(\rho)$ has a constant sign for $c\leq \rho<\infty$ and because $A(\rho)>0$ at spatial infinity so it is for $c\leq \rho<\infty$. Only for P-L solutions, the functions $A(\rho)$, $V(\rho)$ \& $\phi(\rho)$ are finite everywhere for $c\leq \rho<\infty$.

Some static, spherically symmetric, asymptotically flat solutions to Eqs~\eqref{f1} to~\eqref{f4} have been derived in~\cite{B07,N08,B01S}, some of which represent different configurations of \textit{singular} BHs. In Ref~\cite{B01S}, Bronnikov and Shikin constructed BH, NS and P-L solutions. For the latter solution, their function $A$ (Eq (B.3) of Ref~\cite{B01S}) does not approach 1 at spatial infinity ($\rho \to \infty$); rather, it approaches the limit $1+2(\tanh c-1)\text{sech}^2c$, where $c$ is their parametric constant. Thus, their ``P-L" solution is asymptotically flat only in the limit $c \to \infty$, corresponding to the trivial case $r=\rho$ with zero mass and constant functions $A=1,\;\phi=\sqrt{3}\pi/\sqrt{8}\;\&\;V=0$: this is Minkowski space time. Recently, Bronnikov and Chernakova~\cite{B07} constructed a two-parameter asymptotically flat BH characterized by its negative potential $V$ and where $V\;\&\;\phi$ diverge near the center and converge as $\rho\to\infty$, while $A$ is finite everywhere. Very recently, Nikonov \textit{et al}~\cite{N08} have obtained a two-parameter BHs and NSs as well as a one-parameter P-L solutions. For the latter, all the three functions $A>0,\;\phi\;\&\;V$ are finite everywhere, $V$ has both signs and $\phi$ is an incomplete elliptic integral of the first kind.

Most of the solutions constructed previously~\cite{B07,B06,N08,B01S} were derived using the inverse problem method. The method consists in choosing a suitable function $r(\rho)$ then solving Eqs~\eqref{f2} to~\eqref{f4} for $A,\;V\;\&\;\phi$ (Eq~\eqref{f1} is automatically satisfied). There are physically many interesting exact solutions to Eqs~\eqref{f1} to~\eqref{f4} depending on the particular form of $r(\rho)$. The behavior of the solutions at spatial infinity is somehow defined by the boundary condition requiring that the solutions be asymptotically flat
\begin{equation}\label{b}
    A(\rho)=1 - 2m/\rho + O(1/\rho^2)\,,\,r\simeq \rho\;(\rho\to\infty)\;\; \text{and}\;\; \lim_{\rho\to\infty}r'=1\,,
    \,\lim_{\rho\to\infty}r''=0\,,
\end{equation}
where $m$ is the Schwarzschild mass. At the center there is no specific boundary condition, this is why there is a \textit{variety} of possible asymptotically flat solutions, however, grouped into three categories: different (singular) BH solutions, NS and P-L solutions. To be more specific, the BH set ($A$ having both signs) include solutions where $V$ is finite everywhere (this paper) as well as solutions with divergent $V$~\cite{B07}; the same applies to the NS set where $A>0$ (this paper). The solutions discussed in~\cite{B07,N08,B01S} and those discussed in this paper have different boundary conditions at the center.

It is an interesting issue to investigate the behavior of the generic solutions near the center to determine the criteria by which the solutions emerge. The solutions discussed in~\cite{B07,N08,B01S} have been derived by choosing a special function\footnote{The method developed in~\cite{N08} for obtaining the solutions to Eqs~\eqref{f1} to~\eqref{f4} looks somehow different from that developed in~\cite{B07,B01S}, however, it is equivalent. In fact, in~\cite{N08} the function $r(\rho)$ has been chosen such that $\text{d}r/\text{d}\rho=(r^4+4)/(r^4+4-4a^4)$, where $a$ is their parametric constant.} $r(\rho)$, however, they lacked a systematic treatment and many questions remained unanswered. For instance, we do not know a) whether it is possible to obtain a BH (resp. NS) where both $V$ \& $\phi$ are finite everywhere, b) what conditions are required for obtaining, say, a BH or a P-L solution, c) under which conditions the potential $V$ has both signs, etc.

In section~\ref{Sec2}, we derive new integral formulas for $A,\;V\;\&\;\phi$ and proof five corollaries and a theorem on selection criteria for obtaining predefined static, spherically symmetric, asymptotically flat solutions to Eqs~\eqref{f1} to~\eqref{f4}. First we deal with the most generic case then focus on a constrained case with minimum conditions. In section~\ref{Sec3}, the behaviors of $A,\;V\;\&\;\phi$ near the center and at spatial infinity are discussed, instances of solutions of particular interest are provided and two energy conditions are analyzed leading to two further corollaries. In section~\ref{Sec4}, we derive further solutions by the inverse problem method. Two appendices have been added for completeness.

\section{Criteria for obtaining predefined asymptotically flat solutions}\label{Sec2}
Eq~\eqref{f4} is readily brought to the form $[A(r^2)' - r^2A']' = 2$ which is integrated by
\begin{equation}\label{A3}
    r^2A' - (r^2)'A = -2\rho +K\,,
\end{equation}
where $K$ is constant of integration. Integrating~\eqref{A3} leads to
\begin{equation*}
    A(\rho) = Cr^2(\rho) +\left(\int^{\rho}\frac{K-2z}{r^4(z)}
    \,\text{d}z\right)r^2(\rho)\,,
\end{equation*}
where $C$ is a second constant of integration. With the boundary conditions~\eqref{b}, we must have $C=0$ and $K=6m$. Hence,
\begin{equation}\label{A5}
    A(\rho) = 2\left(\int_{\infty}^{\rho}\frac{3m-z}{r^4(z)}
    \,\text{d}z\right)r^2(\rho)\,.
\end{equation}
Using $[A(r^2)' - r^2A']' = 2$ into~\eqref{f2}, we bring $V$ to the form
\begin{equation}\label{V2}
    V(\rho) = -\frac{(r^2A')'}{2r^2}=\frac{1}{r^2} - \frac{(Arr')'}{r^2}\,.
\end{equation}
Substituting~\eqref{A5} into the one or the other expression of $V$, one obtains
\begin{equation}\label{V4}
    V(\rho) = \frac{1}{r^2} - \frac{2(3m-\rho)r'}{r^3} - \frac{A}{r^2}\,(3r'^{\,2}+rr'')\,.
\end{equation}
Finally, without loss of generality we take the negative sign in~\eqref{f3} ($\phi'=-\sqrt{-2r''/r}$) and introduce two arbitrary constants $\rho_0>c$ \& $\phi(\rho_0)$, we obtain
\begin{equation}\label{P2}
    \phi(\rho)-\phi(\rho_0) = -\int_{\rho_0}^{\rho} \sqrt{-2r''(z)/r(z)}\,\text{d}z\,.
\end{equation}

Throughout the remaining part of the paper, when the limits of $A,\;V\;\&\;\phi$ at the center $c$ exist, we define $A(c),\;V(c)\;\&\;\phi(c)$ by $\lim_{\rho\to c^+}A(\rho)$ $\lim_{\rho\to c^+}V(\rho)$ \& $\lim_{\rho\to c^+}\phi(\rho)$, respectively. This definition extends to their derivatives at the center.

\subsection{Generic case}
Notice that Eqs~\eqref{A5}, \eqref{V4} \& \eqref{P2} have been derived without referring to any extra condition (except the boundary conditions~\eqref{b} and, in order to integrate~\eqref{A3}, we have assumed implicitly that $r'(\rho)$ exists ($r(\rho)$ differentiable)), thus the following first two corollaries are generic.

Consider the case where $c\geq 3m$. Since $3m\leq c\leq \rho <\infty$ we have $3m-\rho \leq 0$, thus the integral in~\eqref{A5} is positive leading to $A>0$. Hence,

\begin{corollary}
If the mass of the spherically symmetric, asymptotically flat solution to Eqs~\eqref{f1} to \eqref{f4} is such that $c\geq 3m$ where $c$ is the scaling constant, then the configuration is either a NS or a P-L solution. A negative-masse, spherically symmetric, asymptotically flat solution is either a NS or a P-L solution.
\end{corollary}

Now, let us look for the conditions that make (the limit of) $A$ finite at $c$. Writing $\int_{\infty}^\rho=\int_{\infty}^{u}+\int_{u}^\rho$ in~\eqref{A5}, where $u$ is any fixed number in the open domain ($c,\infty$). The improper integral $\int_{\infty}^{u}\frac{3m-z}{r^4(z)}\,\text{d}z$ converges by the limit comparison test. In fact, let $f(\rho)=|3m-\rho|/r^4$ \& $h(\rho)=1/\rho^3$, since $\lim_{\rho\to\infty}f/h=1$ and since $\int_{\infty}^{u}h(z)\,\text{d}z$ converges for $c<u<\infty$, then the integral of $f$ is convergent. With that said, the value of $A$ at $c$ is the following limit
\begin{eqnarray}
\label{A6}  \lim_{\rho\to c^+}A(\rho) &=& 2\lim_{\rho\to c^+}\frac{\int_{u}^\rho\frac{3m-z}{r^4(z)}\,\text{d}z}{1/r^2(\rho)}\to \frac{\infty}{\infty}\,, \\
\label{A7}   &=& -\frac{1}{r'(c)}\,\lim_{\rho\to c^+}\frac{3m-\rho}{r(\rho)}\,.
\end{eqnarray}
The conditions $r''\leq 0$ (Eq~\eqref{f3}) and $\lim_{\rho\to\infty}r'=1$ (Eq~\eqref{b}) imply $1< r'(\rho) <\infty$ for $c\leq \rho<\infty$. Now, with $r(c)=0$, the limit in~\eqref{A7} exists only\footnote{An instance where $A(c)$ is finite with $(3m-c)$ arbitrary is $r =\sqrt{\rho^2 - c^2}$. However, for this function $r'$ \& $r''$ are divergent at $c$, consequently $V$ too diverges at $c$, thus there is no P-L solution.} if $\lim_{\rho\to c^+}(3m-\rho)=3m-c=0$. In this case $A(c)=\lim_{\rho\to c^+}A(\rho)=1/r'^{\,2}(c)$. If $c>3m$, $\lim_{\rho \to c^+}A(\rho)= +\infty$ and the solution is a NS by corollary 1. If $c<3m$, $\lim_{\rho \to c^+}A(\rho)= -\infty$ and the solution is a BH since $A(\rho)$ has both signs. In this case ($c<3m$), $A(\rho)$ is manifestly positive for $3m\leq \rho<\infty$. For $c\leq \rho<3m$, we can write the integral in~\eqref{A5} as $\int_{\infty}^\rho=\int_{\infty}^{3m}+\int_{3m}^\rho$ where $\int_{\infty}^{3m}\cdots=\text{constant}>0$ and $\int_{3m}^\rho\cdots <0$. As $\rho$ decreases from $3m$ to $c$, the second integral $\int_{3m}^\rho\cdots$ decreases monotonically from $0$ to $-\infty$ (since we found $\lim_{\rho \to c^+}A(\rho)= -\infty$). Consequently, the sum of the two integrals and $A(\rho)$ vanish at one and only one point $\rho_{\text{h}}$ (event horizon) between $c$ and $3m$: $c<\rho_{\text{h}}<3m$. Thus for $c<3m$, $\rho_{\text{h}}$ is a simple root of $A(\rho)=0$ (all horizons are simple).

\begin{corollary}
If $r(\rho)$ is differentiable at $c$, a necessary condition to find a spherically symmetric, asymptotically flat P-L solution to Eqs~\eqref{f1} to \eqref{f4} is $c=3m$ and in this case $A(c)=1/r'^{\,2}(c)$. In the other cases, a) $c>3m$ including negative masses, the solution is a spherically symmetric, asymptotically flat NS with $\lim_{\rho \to c^+}A(\rho)= +\infty$ or b) $c<3m$, the solution is a spherically symmetric, asymptotically flat BH with $\lim_{\rho \to c^+}A(\rho)= -\infty$ and the event horizon $c<\rho_{\rm h}<3m$.
\end{corollary}

The condition $c=3m$ in the above statement for obtaining P-L solutions being necessary, one may derive solutions with $c=3m$ which are not P-L solutions. By corollary 1, these solutions must be NSs. Thus, corollary 2 does not rule out NSs with $c=3m$. Solutions with $c=3m$ and $\lim_{\rho \to c^+}V(\rho)$ divergent are indeed NSs.

We shall see in the next subsection that, for the conditions on which corollary 2 rests, $A(\rho)$ is finite for $0<\rho<\infty$. Since the condition $c=3m$ ensures finiteness of $A$ at $c$, thus by corollary 2 we have solved the eigenvalue problem consisting in determining an everywhere finite solution $A(\rho)$ to Eq~\eqref{f4} with the ``eigenvalue" $m=c/3$.

Repeating the steps~\eqref{A6} \& \eqref{A7} for $\lim_{\rho\to c^+}(rA)$, we obtain
\begin{equation}\label{ben1}
    \hspace{-0.67mm}\lim_{\rho\to c^+}[r(\rho)A(\rho)] = 2\lim_{\rho\to c^+}\frac{\int_{u}^\rho\frac{3m-z}{r^4(z)}\,\text{d}z}{1/r^3(\rho)}=
    -\frac{2}{3r'(c)}\,\lim_{\rho\to c^+}(3m-\rho)=\frac{2(c-3m)}{3r'(c)}\,,
\end{equation}
Hence for $c\neq 3m$, $A$ behaves near the center as
\begin{equation}\label{BA}
    A(\rho) \simeq \frac{2(c-3m)}{3r'(c)}\,\frac{1}{r(\rho)}\,,\;(\rho \to c^+)\,.
\end{equation}

\subsection{Constrained case with minimum conditions}
In addition to the boundary conditions~\eqref{b} and differentiability condition of $r(\rho)$, we impose further (minimum set of) conditions on $r(\rho)$ and its first derivatives in order to derive further selection criteria. We will do that without losing much generality. Thus, throughout the rest of this section, we assume that $r(\rho)$ is four times differentiable in the domain $c\leq \rho <\infty$, which means that there $r(\rho)$ and the derivatives $r'(\rho)$ to $r^{(4)}(\rho)$ are smooth functions (continuous) and finite
\begin{eqnarray}
\label{subj1}& &r(\rho),\;r'(\rho)\;\to\;r^{(4)}(\rho):\;\text{exist\;and\;continuous \;on}\;c\leq \rho <\infty  \\
\label{subj3}& &0\leq r(\rho) <\infty ,\; 1< r'(\rho) <\infty ,\;-\infty < r''(\rho) \leq 0 \,.
\end{eqnarray}
Notice that $r'(\rho)>1$; particularly, we have $1<r'(c)<\infty$. In Part 1 of Appendix A, however, we will see that the third and fourth derivatives of $r(\rho)$ need not exist on the whole domain $\rho\in[c,\infty)$; rather, on a small half-open interval containing $c$.

The conditions~\eqref{subj1} \& \eqref{subj3} do not ensure existence of Taylor series expansions or others for $r(\rho)$ at the center $c$ since the fifth and higher order derivatives of $r(\rho)$ are not assumed to exist there and elsewhere. The same is true for $A(\rho),\;V(\rho)\;\&\;\phi(\rho)$, which are related to $r(\rho)$ via Eqs~\eqref{A5}, \eqref{V4} \& \eqref{P2}. For this reason the method we are considering in this subsection (as was the case in the previous one) and Appendix A is more general, different from and even advantageous than series expansion methods.

We start by proving the finiteness of $V(\rho)\;\&\;\phi(\rho)$ at spatial infinity (that of $A(\rho)$ is imposed in~\eqref{b}). Using~\eqref{b} in~\eqref{V4} \& \eqref{P2} one obtains $\lim_{\rho\to\infty}V=0$ \& $\lim_{\rho\to\infty}\phi=\text{constant}$. The result for $V$ is obvious and to derive that for $\phi$ we take the limit as $\rho\to\infty$ of both sides of~\eqref{P2}
\begin{equation}\label{wr2}
    \lim_{\rho\to\infty}\phi(\rho) = \phi(\rho_0) - \lim_{\rho\to\infty}\int_{\rho_0}^{\rho} \sqrt{-2r''(z)/r(z)}\,\text{d}z\,.
\end{equation}
We can always bring $r(\rho)$ to the form $r(\rho)=\rho-s(\rho)$ such that $\lim_{\rho\to\infty}s=0$, $\lim_{\rho\to\infty}s'=0$ \& $\lim_{\rho\to\infty}s''=0$. More precisely, as $\rho\to\infty$, $s\lesssim a^2/\rho^n$ where $n$ is some positive number\footnote{In the worst case, $s(\rho)$ behaves as $\ln^{\kappa}\rho/\rho$ ($\kappa >0$). Thus as $\rho\to\infty$, $s<1/\sqrt{\rho}$ ($n=1/2$).}. Hence, as $\rho\to\infty$, $\sqrt{-2r''(z)/r(z)}\approx a\sqrt{2n(n+1)}/z^{(n+3)/2}$ and the integral in the r.h.s of~\eqref{wr2} converges to some constant.

For $c<\rho <\infty$ ($0<r <\infty$), the improper integral in~\eqref{A5} converges by the limit comparison test, which means that $A$ is finite and continuous for $c<\rho <\infty$. $A(\rho)$, $r'(\rho)$ \& $r''(\rho)$ being finite functions in the domain $c<\rho <\infty$, Eq~\eqref{V4} implies that $V$ remains finite for $c<\rho <\infty$. Finally, since the function $\sqrt{-2r''(\rho)/r(\rho)}$ is continuous on any closed interval included in the open domain $(c,\infty)$, the integral in~\eqref{P2} exists for $\rho_0>c$, which means that $\phi$ is finite and continuous for $c<\rho <\infty$. We have thus shown that the three functions $A,\;V\;\&\;\phi$ are finite and continuous for $c<\rho <\infty$; the unique point where the functions may diverge is the center $c$.

Notice that the continuity and regularity of the functions $A,\;V\;\&\;\phi$ for $c<\rho <\infty$ is independent of the nature of the solution (BH, NS or P-L solution). Thus

\begin{corollary}
If $r(\rho)$ is constrained by~\eqref{subj1} \& \eqref{subj3}, no singularity at finite $r$ (resp. $\rho$) occurs in the open domain $(0,\infty)$ (resp. $(c,\infty)$) whatever the nature of the solution is.
\end{corollary}

Now, we look for the conditions that make the (limits of the) functions $V\;\&\;\phi$ finite at $c$ (that of $A$ has been established in corollary 2). The behavior of $V$ near the center is derived from~\eqref{V4} as follows.
\begin{eqnarray}
\label{ben2}\lim_{\rho\to c^+}[r^2(\rho)V(\rho)] &=& 1 - \lim_{\rho\to c^+}[(rA)r''] + \lim_{\rho\to c^+}[r'\,T(\rho)] \\
\label{ben3}&=& 1 - \frac{2(c-3m)r''(c)}{3r'(c)} + r'(c)\lim_{\rho\to c^+}T(\rho)\,,
\end{eqnarray}
where we have used~\eqref{ben1} to evaluate the first limit in~\eqref{ben2}. In~\eqref{ben3}, $T(\rho)$ is defined by
\begin{equation*}
    T(\rho) \equiv \frac{2(\rho-3m)-3Ar'r}{r}\;(=-A'r-Ar')\,,
\end{equation*}
which is by~\eqref{ben1} an indeterminate form 0/0 as $\rho\to c^+$. Using Eq~\eqref{A3} with $K=6m$, it is straightforward to show that $T=-(A'r+Ar')$. Thus
\begin{eqnarray}
  \lim_{\rho\to c^+}T &=& \lim_{\rho\to c^+}\frac{2(\rho-3m)-3Ar'r}{r}\nonumber \\
   &=& \frac{2}{r'(c)} - 3\lim_{\rho\to c^+}(A'r+Ar') - \lim_{\rho\to c^+}\frac{3Arr''}{r'}\nonumber\\
\label{ben4} &=& \frac{2}{r'(c)} + 3\lim_{\rho\to c^+}T - \frac{2(c-3m)r''(c)}{r'^{\,2}(c)}\,.
\end{eqnarray}
Solving~\eqref{ben4} for $\lim_{\rho\to c^+}T$ then substituting back into~\eqref{ben3} yields $\lim_{\rho\to c^+}(r^2V)=(c-3m)r''(c)/(3r'(c))$. Thus, for $c\neq 3m$ we have
\begin{equation*}
    V(\rho) \simeq \frac{(c-3m)r''(c)}{3r'(c)}\,\frac{1}{r^2(\rho)}\,,\;(\rho \to c^+)\,.
\end{equation*}

\begin{corollary}
If $r(\rho)$ is two times differentiable with $1<r'(c)<\infty$ ($r(\rho)$ need not satisfy all the conditions~\eqref{subj1} \& \eqref{subj3}) and if $r''(c)\neq 0$, then as $\rho\to c^+$ a) the function $V$ (resp. $A$) for a NS solution with $c>3m$ (corollary 2) approaches $-\infty$ as $-1/r^2$ (resp. $+\infty$ as $1/r$) and b) the function $V$ (resp. $A$) for a BH solution ($c<3m$ by corollary 2) approaches $+\infty$ as $1/r^2$ (resp. $-\infty$ as $-1/r$).
\end{corollary}

Concerning extreme NSs with $c=3m$, we have the following result.

\begin{corollary}
If $r(\rho)$ is two times differentiable with $1<r'(c)<\infty$ ($r(\rho)$ need not satisfy all the conditions~\eqref{subj1} \& \eqref{subj3}) and if $r''(c)\neq 0$, then as $\rho\to c^+$ the function $V$ (resp. $A$) for the NS with $c=3m$ approaches $-\infty$ as $-1/r$ (resp. converges to $1/r'^{\,2}(c)$ by corollary 2).
\end{corollary}

Now, we will look for conditions to have a P-L solution and provide a proof for corollary 5 (Eqs~\eqref{A10}, \eqref{A12} \& \eqref{V5}). By corollary 2, P-L solutions exist only if $c=3m$.

Let $c=3m$. In Part 1 of Appendix A, we will prove the general result that if $c=3m$ then $\lim_{\rho \to c^+}A'$ converges. With $K=6m=2c$, Eq~\eqref{A3} reads $A'(\rho) = [2r'rA+2(c-\rho)]/{r^2}$. Taking the limit of the previous formula as $\rho \to c^+$, replacing $\lim_{\rho \to c^+}A'$ by $A'(c)$ and using $\lim_{\rho \to c^+}A(\rho)=A(c)=1/r'^{\,2}(c)$ we obtain
\begin{align}
& A'(c) = \lim_{\rho \to c^+}\frac{2r'rA+2(c-\rho)}{r^2} \to \frac{0}{0}\nonumber \\
& A'(c) = \lim_{\rho \to c^+}\frac{r''rA+r'rA'+(r'^{\,2}A-1)}{rr'}\nonumber \\
\label{A8}& A'(c) = \frac{r''(c)}{r'^{\,3}(c)} + A'(c) + \frac{1}{r'(c)}\,\lim_{\rho \to c^+}\frac{(r'^{\,2}A-1)}{r}\,.
\end{align}
Dropping $A'(c)$ from both sides of~\eqref{A8}, the remaining equation reads
\begin{eqnarray}
\label{A9}  - \frac{r''(c)}{r'^{\,2}(c)} &=& \lim_{\rho \to c^+}\frac{(r'^{\,2}A-1)}{r}\to \frac{0}{0}\\
- \frac{r''(c)}{r'^{\,2}(c)} &=&  \lim_{\rho \to c^+}\frac{2r'r''A+r'^{\,2}A'}{r'} = \frac{2r''(c)}{r'^{\,2}(c)} + r'(c)\lim_{\rho \to c^+}A'\,.\nonumber
\end{eqnarray}
Solving the last equation for $A'(c)$ we obtain
\begin{equation}\label{A10}
    A'(c)=\lim_{\rho \to c^+}A'(\rho) = -\frac{3r''(c)}{r'^{\,3}(c)}\,.
\end{equation}

Eq~\eqref{f4} reads
\begin{equation}\label{A11}
    rA''=2Ar''+2(r'^{\,2}A-1)/r\,.
\end{equation}
Using~\eqref{A9} and $\lim_{\rho \to c^+}A(\rho)=A(c)=1/r'^{\,2}(c)$ in~\eqref{A11} we obtain
\begin{equation}\label{A12}
    \lim_{\rho \to c^+}(rA'')=\frac{2r''(c)}{r'^{\,2}(c)} - 2\,\frac{r''(c)}{r'^{\,2}(c)}=0\,.
\end{equation}

The first equation in~\eqref{V2} reads
\begin{equation}\label{V5}
    V(\rho) = -\frac{2r'A'+rA''}{2r}\,.
\end{equation}
Notice that by~\eqref{A10} \& \eqref{A12}, $\lim_{\rho \to c^+}(2r'A'+rA'')=-6r''(c)/r'^{\,2}(c)$, thus if $r''(c)\neq 0$ ($r''(c)<0$) then $\lim_{\rho \to c^+}$ $V(\rho)$ approaches $-\infty$ as $-1/r$; this establishes a proof for corollary 5. In this case $V$ \& $A$ behave near the center as
\begin{equation*}
    c=3m:\;\;V(\rho) \simeq \frac{3r''(c)}{r'^{\,2}(c)}\,\frac{1}{r(\rho)}\,,\;A(\rho)\simeq \dfrac{1}{r'^{\,2}(c)}\,,\;(\rho \to c^+)\,.
\end{equation*}

Now, we further assume that $r''(c)=0$ (with $c=3m$), the limit of both sides of~\eqref{V5} as $\rho \to c^+$ leads to
\begin{eqnarray}
\label{n1} \lim_{\rho \to c^+}V &=&  -r'(c)\lim_{\rho \to c^+}\frac{A'}{r}-\lim_{\rho \to c^+}\frac{A''}{2}\\
\label{n2} &=& -r'(c)\lim_{\rho \to c^+}\frac{A''}{r'}-\lim_{\rho \to c^+}\frac{A''}{2} = -\frac{3}{2}\,\lim_{\rho \to c^+}A''\,.
\end{eqnarray}
Since $r''(c)=0$ this implies $A'(c)=0$ by~\eqref{A10} making thus the first limit in the r.h.s of~\eqref{n1} an indeterminate form 0/0. Next, consider the second term in~\eqref{V2}
\begin{eqnarray}
  \lim_{\rho \to c^+}V &=& \lim_{\rho \to c^+}\frac{1-A'rr'-Ar'^{\,2}-Arr''}{r^2} \to \frac{0}{0}\nonumber\\
\label{an} &=& -\lim_{\rho \to c^+}\left(r'\frac{A'}{r}+\frac{A'r''}{r'}+\frac{3A}{2}\,\frac{r''}{r}+\frac{Ar'''}{2r'}+\frac{A''}{2}\right)\,,
\end{eqnarray}
where the second limit in~\eqref{an} vanishes by the imposed condition: $r''(c)=0$. The first and third limits are indeterminate forms 0/0, leading to upon further differentiating
\begin{equation}\label{n3}
    \lim_{\rho \to c^+}V = -\frac{3}{2}\,\lim_{\rho \to c^+}A'' - \frac{2r'''(c)}{r'^{\,3}(c)}\,.
\end{equation}
Eqs~\eqref{n2} \& \eqref{n3} result in
\begin{equation}\label{le}
    \lim_{\rho \to c^+}V = -\frac{3}{2}\,\lim_{\rho \to c^+}A''\,\;\;\&\;\;r'''(c)=0\,.
\end{equation}

In Part 1 of Appendix A we will prove the general result that if $c=3m$ \& $15r''^{\,2}(c)-4r'(c)r'''(c)=0$ then $\lim_{\rho \to c^+}A''$ ($=A''(c)$) converges and provide a formula for $A''(c)$. In our case, since $r''(c)=0$ \& $r'''(c)=0$, the latter condition is satisfied and the desired formula for $A''(c)$ will have the form (see Eq~\eqref{ap15})
\begin{equation}\label{new}
    A''(c)=2\lim_{\rho \to c^+}(r'^{\,2}A-1)/r^2\equiv L\,,\;\text{if}\;c=3m\,,\;r''(c)=0\,,\;\&\; r'''(c)=0\,.
\end{equation}
Notice that~\eqref{new} is derived from~\eqref{A11} after dropping the first term, which is zero: $\lim_{\rho \to c^+}r''/r=\lim_{\rho \to c^+}r'''/r'=0$.

Eq~\eqref{P2} implies $\phi'(c)=\lim_{\rho \to c^+}\phi'$ $=$ $-\lim_{\rho \to c^+}\sqrt{-2r''/r}$ $=$ $-[\lim_{\rho \to c^+}$ \\$(-2r''/r)]^{1/2}$ $=$ $-[\lim_{\rho \to c^+}(-2r'''/r')]^{1/2}=0$ (by~\eqref{le}), thus $\phi$ is differentiable at $c$ and consequently it is continuous there, which means that $\phi(c)$ exists and is finite. Finally, we managed to make the three function $A,\;V\;\&\;\phi$ finite at $c$, thus finite everywhere in the domain $c\leq \rho <\infty$ under the following conditions.

\begin{theorem}
Suppose $r(\rho)$ is four times differentiable at every point of the domain $c\leq \rho <\infty$ and satisfies the conditions~\eqref{subj1} \& \eqref{subj3} ($r'''$ \& $r^{(4)}$ need not exist on the whole domain $\rho\in[c,\infty)$; rather, on a small half-open interval containing $c$). If $r''(c)=0$, $r'''(c)=0$ \& $c=3m$, any spherically symmetric, asymptotically flat solution to Eqs~\eqref{f1} to \eqref{f4} is a one parameter P-L configuration with $A(c)=1/r'^{\,2}(c)$, $A'(c)=0$ \& $V(c)=-3A''(c)/2$.
\end{theorem}

Table~\ref{Tab1} summarizes the results of the above discussions and includes some results from the following section.

\begin{table}
\caption{{\small Static, spherically symmetric, asymptotically flat solutions to Eqs~\eqref{f1} to~\eqref{f4}. Column 1 from the left shows the statements concerning the behavior of the derivatives of $r(\rho)$ and their values at the center $\rho=c$ ($r=0$). At spatial infinity $\rho=\infty$, the boundary conditions~\eqref{b} are assumed for all solutions. In Columns 2, 3 \& 4 the behaviors of $A(\rho),\;V(\rho)\;\&\;\phi(\rho)$, when available, are given in the limit $\rho\rightarrow c^+$ ($r\rightarrow 0^+$) and only the dominant terms are shown. Notation \& nomenclature: $\Delta\phi\equiv \phi(\rho)-\phi(c)$, $x\equiv \rho-c$, ``NC" for ``Necessary Condition", ``SC" for ``Sufficient Condition". In the expressions of $V$ depending on $\ln x$, we assume existence of $r^{(5)}(\rho)$ as stated in Eqs~\eqref{s2} to~\eqref{s2b}.}}
\label{Tab1}

{\small
\tablefirsthead{
  \noalign{\smallskip}
  \hline
  \multicolumn{1}{|c|}{\textsc{Condition(s):}}  &
  \multicolumn{1}{c|}{\textsc{Particle-Like Solution}}  &
  \multicolumn{1}{c|}{\textsc{Naked Singularity}}  &
  \multicolumn{1}{c|}{\textsc{Black Hole}}  \\
  \multicolumn{1}{|c|}{At $\rho=\infty$ the conditions}  &
  \multicolumn{1}{c|}{}  &
  \multicolumn{1}{c|}{}  &
  \multicolumn{1}{c|}{}  \\
  \multicolumn{1}{|c|}{\eqref{b} are assumed}  &
  \multicolumn{1}{c|}{$A(\rho)>0$}  &
  \multicolumn{1}{c|}{$A(\rho)>0$}  &
  \multicolumn{1}{c|}{$A(\rho)$ has both signs}  \\
  \hline}
\tablehead{
  \hline
  \multicolumn{1}{|c|}{\textsc{Condition(s):}}  &
  \multicolumn{1}{c|}{\textsc{Particle-Like Solution}}  &
  \multicolumn{1}{c|}{\textsc{Naked Singularity}}  &
  \multicolumn{1}{c|}{\textsc{Black Hole}}  \\
  \multicolumn{1}{|c|}{At $\rho=\infty$ the conditions}  &
  \multicolumn{1}{c|}{}  &
  \multicolumn{1}{c|}{}  &
  \multicolumn{1}{c|}{}  \\
  \multicolumn{1}{|c|}{\eqref{b} are assumed}  &
  \multicolumn{1}{c|}{$A(\rho)>0$}  &
  \multicolumn{1}{c|}{$A(\rho)>0$}  &
  \multicolumn{1}{c|}{$A(\rho)$ has both signs}  \\
  \hline}
\tabletail{
  \hline
  \multicolumn{4}{r}{\textsl{Continued on next page}}\\
  \hline}
\tablelasttail{\hline}
\begin{supertabular*}{1.011\textwidth}{|l|l|l|l|}
\textsc{Corollary 2:} & $c=3m$ (NC): & a) $m<c/3$ (SC): & $m>c/3$ (SC): \\
$r'(\rho)$ exists & $A(\rho)\simeq \dfrac{1}{r'^{\,2}(c)}$ & $A(\rho)\simeq \dfrac{2(c-3m)}{3r'(c)}\dfrac{1}{r(\rho)}$ & $A(\rho)\simeq \dfrac{2(c-3m)}{3r'(c)}\dfrac{1}{r(\rho)}$ \\
 &  & b) $c=3m$: Possible &  \\
\hline
\textsc{Corollaries 4 \& 5:} & & a) $m<c/3$ (SC): & $m>c/3$ (SC): \\
$r'(\rho)$ \& $r''(\rho)$ exist, & & $A(\rho)\simeq \dfrac{2(c-3m)}{3r'(c)}\dfrac{1}{r(\rho)}$ & $A(\rho)\simeq \dfrac{2(c-3m)}{3r'(c)}\dfrac{1}{r(\rho)}$ \\
 $r''(c)\neq 0$ & & $V(\rho) \simeq \dfrac{(c-3m)r''(c)}{3r'(c)}\dfrac{1}{r^2(\rho)}$ & $V(\rho) \simeq \dfrac{(c-3m)r''(c)}{3r'(c)}\dfrac{1}{r^2(\rho)}$ \\
 & & b) $c=3m$: & \\
 & & $A(\rho)\simeq \dfrac{1}{r'^{\,2}(c)}$ &\\
  & & $V(\rho)\simeq \dfrac{3r''(c)}{r'^{\,2}(c)}\dfrac{1}{r(\rho)}$ & \\
\hline
\textsc{Theorem:} & $c=3m$ (SC): & $m<c/3$ (SC): & $m>c/3$ (SC): \\
$r'(\rho)$, $r''(\rho)$, $r'''(\rho)$ & $A(\rho)\simeq \dfrac{1}{r'^{\,2}(c)}$ \& $A'(c)=0$ & $A(\rho)\simeq \dfrac{2(c-3m)}{3r'(c)}\dfrac{1}{r(\rho)}$ & $A(\rho)\simeq \dfrac{2(c-3m)}{3r'(c)}\dfrac{1}{r(\rho)}$ \\
\& $r^{(4)}(\rho)$ exist, & $V(\rho)\simeq \dfrac{-3L}{2}$ & $V(\rho) \simeq \dfrac{(3m-c)r^{(4)}(c)}{r'^{\,3}(c)}\,\ln x$ & $V(\rho) \simeq \dfrac{(3m-c)r^{(4)}(c)}{r'^{\,3}(c)}\,\ln x$ \\
$r''(c)=0$, $r'''(c)=0$ & $\Delta\phi \simeq - \dfrac{2}{3}\sqrt{\dfrac{-r^{(4)}(c)}{r'(c)}}\,x^{3/2}$ & $\Delta\phi \simeq - \dfrac{2}{3}\sqrt{\dfrac{-r^{(4)}(c)}{r'(c)}}\,x^{3/2}$ & $\Delta\phi \simeq - \dfrac{2}{3}\sqrt{\dfrac{-r^{(4)}(c)}{r'(c)}}\,x^{3/2}$ \\
\end{supertabular*}}
\end{table}

Most of the solutions constructed previously~\cite{B07,B06,N08,B01S} were derived using the inverse problem method, which we are aiming to use in section~\ref{Sec4}. The method consists in choosing a suitable function $r(\rho)$ then using Eqs~\eqref{A5}, \eqref{V4} \& \eqref{P2} to obtain $A,\;V\;\&\;\phi$.

From the above discussions (five corollaries and a theorem), we have learnt how to set conditions on $r(\rho)$ and to fix the sign of $(c-3m)$ in order to obtain a predefined final solution. For instance, for $r(\rho)$ given by
\begin{equation}\label{R1}
    r(\rho) = \frac{n}{n-1}\,(\rho - c) - \frac{1}{n-1}\,\frac{(\rho - c)^n}{\rho^{n-1}}\,,\;\rho \geq c>0\,,\;n\;\text{(integer)}\geq 2\,,
\end{equation}
(where $r''(\rho)=-nc^2(\rho-c)^{n-2}/\rho^{n+1}\leq 0$) if we choose $n\geq 4$, then $r(\rho)$ satisfies all the requirements of the theorem but one; if we further choose $c=3m$ in~\eqref{A5}, then Eqs~\eqref{A5}, \eqref{V4} \& \eqref{P2} lead to a one-parameter P-L solution for each value of $n\geq 4$. If, instead, we choose $c>3m$ or $c<3m$ the solution is by corollary 2 a two-parameter NS or a BH, respectively, for each value of $n\geq 4$.

Notice how crucial is the condition $r'''(c)=0$ for obtaining P-L solutions for
the set of the continuous everywhere four times differentiable functions $r(\rho)$ satisfying the conditions~\eqref{subj1} \& \eqref{subj3}. For $n=3$, this condition fails ($r''(c)=0$ \& $r'''(c)=-3/c^4$) and the above formula~\eqref{R1} gives rise to no P-L solution, whatever the value of $(3m-c)$ is; even when $c=3m$, the potential $V$ still diverges at $c$ as $r'''(c)\ln (\rho - c)$ (see Eq~\eqref{v11}). Similar conclusion holds for $n=2$.

In the following section, we will derive further properties of the solutions at both the center and spatial infinity for more restrictive conditions on $r(\rho)$. We will also discuss the null and weak energy conditions.

\section{Generic properties and energy conditions}\label{Sec3}
Here we study the properties of the asymptotically flat solutions (in their generic forms) by investigating their behavior near the center and at spatial infinity. We also examine the weak energy condition (WEC) at these end-points and the null energy condition (NEC) on the whole range of $\rho$.

\subsection{The center vs. spatial infinity}
In this section, we rather assume that $r(\rho)$ is five times differentiable:
\begin{eqnarray}
\label{s2}& &r(\rho),\;r'(\rho)\;\to\;r^{(5)}(\rho):\;\text{exist\;and\;continuous \;on}\;c\leq \rho <\infty  \\
\label{s2b}& &0\leq r(\rho) <\infty ,\; 1< r'(\rho) <\infty ,\;-\infty < r''(\rho) \leq 0 \,.
\end{eqnarray}

The behavior of $A$ near the center is discussed in 1) Part 1 of Appendix A for $r(\rho)$ satisfying~\eqref{subj1} \& \eqref{subj3} with $c=3m$ and 2) Part 2 of Appendix A for $r(\rho)$ satisfying~\eqref{s2} \& \eqref{s2b} with $c\neq 3m$.

If $r(\rho)$ satisfies~\eqref{s2} \& \eqref{s2b}, $A,\;V\;\&\;\phi$ take the following forms near the center $\rho=c$ ($r=0$), where $A''(c)$ is determined in Part 2 of Appendix A by~\eqref{ap16}
\begin{eqnarray}
A &=& \frac{2(c-3m)}{3r'^{\,2}(c)}\,\frac{1}{x} + \frac{1}{r'^{\,2}(c)} - \frac{4(c-3m)r''(c)}{3r'^{\,3}(c)} + \frac{F}{r'^{\,4}(c)}\,x \nonumber \\
\label{a1} & & + \frac{G}{r'^{\,5}}\,x^2\ln x + \frac{A''(c)}{2}\,x^2 +\cdots\,,
\end{eqnarray}
\begin{eqnarray}
V &=& \frac{(c-3m)r''(c)}{3r'^{\,3}(c)}\,\frac{1}{x^2} - \frac{10(c-3m)r''^{\,2}(c)-r'(c)[9r''(c)+4(c-3m)r'''(c)]}{3r'^{\,4}(c)}\,\frac{1}{x}\nonumber \\
\label{v11} & & -\frac{3A''(c)}{2}-\frac{14(c-3m)r''^{\,3}(c)}{r'^{\,5}(c)} - \frac{10r'''(c)}{3r'^{\,3}(c)} +
   \frac{157(c-3m)r''(c)r'''(c)}{18r'^{\,4}(c)} \\
 & & - \frac{3(c-3m)r^{(4)}(c)}{4r'^{\,3}(c)} + \frac{14r''^{\,2}(c)}{r'^{\,4}(c)} - \frac{3G}{r'^{\,5}(c)}\ln x + \frac{H}{r'^{\,6}(c)}\,x\ln x+\cdots\,,\nonumber
\end{eqnarray}
where $x\equiv \rho -c$ and $F$, $G$ \& $H$ are polynomials of $(r'(c),r''(c),r'''(c),r^{(4)},(c-3m))$. The polynomials $F$ \& $G$ are given by
\begin{eqnarray*}
  F&=&\frac{57(c-3m)r''^{\,2}(c)-r'(c)[54r''(c)+20(c-3m)r'''(c)]}{18}\,,\\
  G &=& \frac{4r'''(c)r'^{\,2}(c)-15r''^{\,2}(c)r'(c)}{3} + 5(c-3m)r''^{\,3}(c) \\
   & & - \frac{10(c-3m)r''(c)r'''(c)r'(c)}{3} + \frac{(c-3m)r^{(4)}(c)r'^{\,2}(c)}{3}\,.
\end{eqnarray*}

From now on and throughout the rest of this paper, we choose $\rho_0=\infty$ and $\phi(\infty)=0$ (precisely $\lim_{\rho\to\infty}\phi(\rho)\to 0^+$) in~\eqref{P2}. The function $r(\rho)$ being concave down on its range, the first nonzero derivative $r^{(n)}(c)$, $2\leq n\leq 5$, is negative ($r'(c)>1$ always). The behavior of $\phi(\rho)$ depends of the first non-zero $r^{(n)}(c)$:
\begin{eqnarray}
\phi(\rho) &=& \phi(c) - \sqrt{8}\,\sqrt{\frac{-r''(c)}{r'(c)}}\,\sqrt{x} + \cdots\,,\;\text{if}\;r''(c)\neq 0\,,\nonumber \\
\label{p11} \phi(\rho) &=& \phi(c) - \sqrt{2}\,\sqrt{\frac{-r'''(c)}{r'(c)}}\,x + \cdots\,,\;\text{if}\;r''(c)=0,\,r'''(c)\neq 0\,, \\
\phi(\rho) &=& \phi(c) - \frac{2}{3}\,\sqrt{\frac{-r^{(4)}(c)}{r'(c)}}\,x^{3/2} + \cdots\,,\;\text{if}\;r''(c)=0,\,r'''(c)=0,\,r^{(4)}(c)\neq 0\,,\nonumber
\end{eqnarray}
where $\phi(c)>0$ ($\phi'(\rho)<0$ and we have chosen $\lim_{\rho\to\infty}\phi(\rho)\to 0^+$).

The behavior at spatial infinity is somehow defined by the boundary condition~\eqref{b}. This condition~\eqref{b} does not completely fix the behavior of $A,\;V\;\&\;\phi$ at spatial infinity. We fix that by requiring $r(\rho)$ of the form $r(\rho) = \rho + a_0+\sum_{n=1}^\infty(a_n/\rho)$. The constant $a_0$ can be absorbed in the first term upon performing the translation\footnote{Such a translation modifies the value of the new $\rho$ at $r=0$, which is now $c+a_0$, without modifying the scaling constant ($=c$). In Appendix A, we perform a similar translation and work with a radial coordinate $x$ whose value at $r=0$ is $0$ instead of $c$. Such a translation is no trouble, only some equations, like~\eqref{A5}, change the form (compare with~\eqref{nw4b}), however, the derivations and conclusions are the same.} $\rho\rightarrow \rho + a_0$. Thus, at spatial infinity we assume $r(\rho)$ of the form
\begin{equation}\label{s1}
    r(\rho) = \rho + \sum_{n=1}^\infty(a_n/\rho)\,.
\end{equation}
This is the simplest form to which one may bring $r(\rho)$ to. The solutions discussed in~\cite{B07,N08,B01S} have the same expansion as in~\eqref{s1}.

Upon substituting~\eqref{s1} into~\eqref{A5} then into~\eqref{V4}, one obtains
\begin{align}
 \label{A1} & A(\rho)=1 - \frac{2m}{\rho} + \frac{2(2ma_1+a_2)}{5\rho^3} + \frac{a_1^2+2a_3}{3\rho^4} + O(1/\rho^5)\,,\\
 \label{V1} & V(\rho)=\frac{2(4ma_1-3a_2)}{5\rho^5} + \frac{2(3ma_2-a_1^2-2a_3)}{\rho^6} + O(1/\rho^7)\,,
\end{align}

Notice that, since the function $r(\rho)$ is concave down on its range, the first nonzero constant $a_\ell$ in~\eqref{s1} is negative. Accordingly, the series expansions for $\phi$, which are derived from~\eqref{P2} or~\eqref{f3}, read
\begin{eqnarray*}
  \phi(\rho) &=& \frac{2\sqrt{-a_1}}{\rho} - \frac{3a_2}{2\sqrt{-a_1}\rho^2} + O(1/\rho^3)\,,\;\text{if}\;a_1\neq 0\,, \\
 \phi(\rho) &=& \frac{4\sqrt{-a_2}}{\sqrt{3}\rho^{3/2}} - \frac{4\sqrt{3}a_3}{5\sqrt{-a_2}\rho^{5/2}} + O(1/\rho^{7/2})\,,\;\text{if}\;a_1=0,\,a_2\neq 0\,, \\
  \phi(\rho) &=& \frac{\sqrt{-6a_3}}{\rho^2} - \frac{5\sqrt{2}a_4}{3\sqrt{3}\sqrt{-a_3}\rho^3} + O(1/\rho^4)\,,\;\text{if}\;a_1=a_2=0,\,a_3\neq 0\,.
\end{eqnarray*}

One draws the following conclusions concerning the behavior of $V,\;\phi\;\&\;A$ as $\rho\to\infty$. Although $V\to 0$ as $\rho\to\infty$, it may approach zero from above or from below by adjusting the parameters $a_n$. For instance if $a_1\neq 0$, $a_2\neq 0$ \& $a_1m>3a_2/4$, the potential is positive at spatial infinity, which means the weak energy condition (WEC) is satisfied. Since in this case $a_1<0$, the previous condition on $m$ is satisfied if $m<3a_2/(4a_1)$: the mass of the solution is bounded from above by $3a_2/(4a_1)$. Particular solutions where the mass is bounded from above have been obtained in~\cite{N08}. Conversely, if the mass is bounded from blow $m>3a_2/(4a_1)$, the potential is negative at spatial infinity. We will show in the following subsection that in this case the WEC is also satisfied. If, otherwise $a_1=a_2=0$ but $a_3\neq 0$, the potential is always positive at spatial infinity ($\forall\;m$) since in this case $a_3<0$ an again the WEC is satisfied. The scalar $\phi$, which also vanishes at spatial infinity, has always the same sign as $\rho\to\infty$. Finally, for $m>0$, $A$ always approaches its limit 1 from below.

\begin{corollary}
At spatial infinity, we assume that $r(\rho)-\rho$ admits an expansion in powers of $1/\rho$ given by~\eqref{s1}. Then if $a_1\neq 0$ ($a_1<0$) \& $a_2\neq 0$, there is a critical mass $m_c=3a_2/(4a_1)$ beyond which ($m>m_c$) the potential is negative at spatial infinity; if, instead, $m<m_c$ the potential is positive. For $m=m_c$ the potential may have any sign. In all cases, the WEC is satisfied at spatial infinity.
\end{corollary}

In the paragraph preceding corollary 3 we have mentioned that the finiteness of $A,\;V\;\&\;\phi$ for $c<\rho <\infty$ is independent of the nature of the solution. A first important and yet general similar result from the set of Eqs~\eqref{a1}, \eqref{v11} \& \eqref{p11} is that $\phi$ is always finite at the center, provided conditions~\eqref{s2} \& \eqref{s2b} are satisfied. Although $A(c)$ is finite for $c=3m$, as seen from Eq~\eqref{a1}, it is obvious from Eq~\eqref{v11} that the condition $c=3m$ alone does not ensure existence of P-L solutions for, in this case, $V$ diverges near the center as $3r''(c)/[r'^{\,3}(c)x]\to -\infty$ and the solution is a NS. Similarly, the conditions $r''(c)=0\;\&\;c=3m$ do not ensure existence of P-L solutions for $V$ diverges near the center as $-4r'''(c)\ln x/r'^{\,3}(c)\to -\infty$ and the solution is again a NS.

Other important conclusions derived from Eqs~\eqref{a1} \& \eqref{v11}, not discussed yet in the literature, follow. Under constraints~\eqref{s2} \& \eqref{s2b}:\\

\n a) The field Eqs~\eqref{f1} to~\eqref{f4} do not admit BH solutions ($c<3m$) where the metric $A$ is finite everywhere. However, one can derive a BH solution where both $V\;\&\;\phi$ are \textit{finite} everywhere by choosing $r''(c)=r'''(c)=r^{(4)}(c)=0$; in this case, $A$ diverges at the center as $2(c-3m)/[3r'^{\,2}(c)x]\to -\infty$ (\eqref{a1} or~\eqref{BA}). One may call such a solution a ``semi-regular" BH since all matter fields are finite. We have a set of such two-parametric solutions given by~\eqref{R1} provided we take $n\geq 5$ to ensure that $r''(c)=r'''(c)=r^{(4)}(c)=0$ and in~\eqref{A5} we choose $c<3m$. The simplest of these solutions is
\begin{equation}\label{R2}
    r(\rho) = \frac{5}{4}\,(\rho - c) - \frac{1}{4}\,\frac{(\rho - c)^5}{\rho^{4}}\,,\;\rho \geq c>0\,,\;c<3m\,.
\end{equation}
In Ref~\cite{B07}, another ``semi-regular" black hole solution with $A$ finite everywhere has been constructed, however, by choosing a function $r(\rho)$ not differentiable at the center for which $\lim_{\rho \to c^+}r'(\rho)=\infty$ (in our solutions $1<r'(c)<\infty$). This choice of $r(\rho)$ caused  $V\;\&\;\phi$ to diverge at the center.\\

\n b) Similarly, if $c>3m$ the field Eqs~\eqref{f1} to~\eqref{f4} do not admit NS solutions where $A$ is finite everywhere. However, if $c>3m$ one can obtain a NS where both $V\;\&\;\phi$ are \textit{finite} everywhere by choosing $r''(c)=r'''(c)=r^{(4)}(c)=0$; in this case, $A$ diverges at the center as $2(c-3m)/[3r'^{\,2}(c)x]\to +\infty$ (\eqref{a1} or~\eqref{BA}). We have a set of such two-parametric solutions given by~\eqref{R1} provided we take $n\geq 5$ to ensure that $r''(c)=r'''(c)=r^{(4)}(c)=0$ and in~\eqref{A5} we choose $c>3m$. The simplest of these solutions is
\begin{equation}\label{R3}
    r(\rho) = \frac{5}{4}\,(\rho - c) - \frac{1}{4}\,\frac{(\rho - c)^5}{\rho^{4}}\,,\;\rho \geq c>3m>0\,.
\end{equation}
\n \\
\n c) For any NS with $c=3m$, $A$ is finite everywhere and $V$ diverges at the center as $-4r'''(c)\ln x/r'^{\,3}(c)\to -\infty$ if ($r''(c)=0\;\&\;r'''(c)\neq 0$) or as $3r''(c)/[r'^{\,3}(c)x]\to -\infty$ if $r''(c)\neq 0$. A solution of the case ($r''(c)=0\;\&\;r'''(c)\neq 0$) is (Eq~\eqref{R1} with $n=3$)
\begin{equation}\label{R4}
    r(\rho) = \frac{3}{2}\,(\rho - c) - \frac{1}{2}\,\frac{(\rho - c)^3}{\rho^{2}}\,,\;\rho \geq c=3m>0\,,
\end{equation}
and a solution of the case $r''(c)\neq 0$ is (Eq~\eqref{R1} with $n=2$) is
\begin{equation}\label{R5}
    r(\rho) = 2(\rho - c) - \frac{(\rho - c)^2}{\rho}\,,\;\rho \geq c=3m>0\,.
\end{equation}

\subsection{Null and weak energy conditions}
For the case of self-gravitating real scalar field, the NEC and WEC are the mostly discussed conditions in the literature~\cite{B01D,beken}. If $k^\mu\;\&\;u^\mu$ are \textit{any} future-directed null ($k_\mu k^\mu=0$) and timelike ($u_\mu u^\mu=1$) four-vectors, the NEC and WEC state
\begin{equation}\label{ec1}
    \text{NEC:}\;\;T^{\mu}{}_{\nu}k_{\mu}k^{\nu} \geq 0\,,\quad \text{WEC:}\;\;T^{\mu}{}_{\nu}u_{\mu}u^{\nu} \geq 0\,.
\end{equation}

It was established in~\cite{B01D} that the conditions which rule out wormholes, horns and flux tubes rest on the validity of the NEC. In this section we will show that (P-L, NS or BH) solutions to Eqs~\eqref{f1} to~\eqref{f4} do not violate the NEC. Hence, the consideration of the NEC does not allow for further selection criteria.

Let $\epsilon$ denotes the sign of $A$ such that $\epsilon^2=1\;\&\;\epsilon A=|A|$ ($\epsilon =-1$ only inside a BH). A tetrad frame associated with the metric~\eqref{m} is given by
\begin{eqnarray}\label{ec2}
\begin{array}{l}
e_{0\mu}=(\epsilon \sqrt{|A|},0,0,0)\,,\;e_{1\mu}=(0,-\epsilon/\sqrt{|A|},0,0)\,,\;
e_{2\mu}=(0,0,-r,0)\,,\\
e_{3\mu}=(0,0,0,-r\sin \theta)\,,\;e_0{}^\mu=(1/\sqrt{|A|},0,0,0)\,,\;e_1{}^\mu=(0,\sqrt{|A|},0,0)\,,\\
e_2{}^\mu=(0,0,1/r,0)\,,\;e_3{}^\mu=(0,0,0,1/(r\sin \theta))\,.
\end{array}
\end{eqnarray}
Future-directed null ($k^\mu$) and four-velocity ($u^\mu$) vectors take the forms
\begin{align}
\label{ec3} & k^\mu = e_0{}^\mu + s_1e_1{}^\mu + s_2e_2{}^\mu + s_3e_3{}^\mu \,,\,\;\text{with}\;\;\epsilon(1-s_1^2)-s_2^2-s_3^2=0\,,\\
\label{ec4} & u^\mu = N(e_0{}^\mu + s_1e_1{}^\mu + s_2e_2{}^\mu + s_3e_3{}^\mu) \,,\,\;\text{with}\;\;N=1/\sqrt{\epsilon(1-s_1^2)-s_2^2-s_3^2}\,.
\end{align}

For the static, spherically symmetric solutions to Eqs~\eqref{f1} to~\eqref{f4}, the fields $A,\;V\;\&\;\phi$ depend only on $\rho$. Eqs~\eqref{m} \& \eqref{em} lead then to $T^t{}_t=T^{\theta}{}_{\theta}=T^{\varphi}{}_{\varphi}=V+(A\phi'^{\,2}/2)$, $T^{\rho}{}_{\rho}=V-(A\phi'^{\,2}/2)$ and $T^{\mu}{}_{\nu}=0$ if $\mu\neq \nu$. Hence
\begin{equation*}
    T^{\mu}{}_{\nu}k_{\mu}k^{\nu}=\epsilon s_1^2 (T^t{}_t - T^{\rho}{}_{\rho}) = s_1^2 |A| \phi'^{\,2}\,.
\end{equation*}
Since $s_1^2 |A| \phi'^{\,2}\geq 0$ (the r.h.s is zero if $s_1=0$ (with $\epsilon =1$) or/and $r''(c)=0$), solutions to Eqs~\eqref{f1} to~\eqref{f4} do not violate the NEC. Notice that the NEC is fulfilled even if the field Eqs~\eqref{f1} to~\eqref{f4} are not satisfied.

Similarly, we obtain
\begin{align}
T^{\mu}{}_{\nu}u_{\mu}u^{\nu}=& N^2[(\epsilon-s_2^2-s_3^2)T^t{}_t - \epsilon s_1^2T^{\rho}{}_{\rho}]\nonumber\\
\label{ec5} = & N^2\{[\epsilon (1-s_1^2)-s_2^2-s_3^2]V + [1+s_1^2-\epsilon s_2^2-\epsilon s_3^2]|A|\phi'^{\,2}/2\}\,.
\end{align}
The number in the first square parenthesis of~\eqref{ec5} is positive by~\eqref{ec4} ($N$ real). This leads to $1+s_1^2-\epsilon s_2^2-\epsilon s_3^2>2s_1^2$ if $\epsilon =1$ or $1+s_1^2-\epsilon s_2^2-\epsilon s_3^2\geq s_1^2+1$ if $\epsilon =-1$. Thus, at the points where $V\geq 0$, the r.h.s of~\eqref{ec5} is positive and the WEC is satisfied.

The sign of the r.h.s of~\eqref{ec5} cannot be determined without knowing the solution to Eqs~\eqref{f1} to~\eqref{f4}. Since the functions $A,\;V\;\&\;\phi'$ are related to $r(\rho)$ and its derivatives via Eqs~\eqref{A5}, \eqref{f2} \& \eqref{f3}, it suffices to know the shape of $r(\rho)$ to completely determine the sign of the r.h.s of~\eqref{ec5}. In this paper we have chosen $r(\rho)$ by imposing the conditions~\eqref{b} \& \eqref{s1} at spatial infinity and the conditions~\eqref{s2} \& \eqref{s2b} at the center.

Consider $r(\rho)$ of the form~\eqref{s1} and let $\rho^\phi\,\&\,\;p^\phi_\rho$ be the mass density and radial pressure due to the field $\phi$. Expressing $A,\;V\;\&\;\phi'$ in powers of $1/\rho$, we obtain
\begin{align}
\label{ec6} & T^{\mu}{}_{\nu}u_{\mu}u^{\nu}=-\frac{2\ell(\ell+1)[(\ell+1)(1-s_2^2-s_3^2)+2s_1^2]a_\ell}{(\ell+3)\rho^{\ell+3}}
+\cdots\,,\;\;(\ell\geq 1)\\
& \rho^\phi = T^{t}{}_{t}=-\frac{2\ell(\ell+1)^2a_\ell}{(\ell+3)\rho^{\ell+3}}
+\cdots\,, \;\; p^\phi_\rho = -T^{\rho}{}_{\rho}=-\frac{4\ell(\ell+1)a_\ell}{(\ell+3)\rho^{\ell+3}}
+\cdots\,,\nonumber
\end{align}
(at spatial infinity $\epsilon =1$ \& $1-s_1^2-s_2^2-s_3^2>0$). Here $a_\ell$ is the first non-zero constant $a_n$ in~\eqref{s1}: $a_\ell<0$. Thus, the dominant term in the r.h.s of~\eqref{ec6} is positive.

\begin{corollary}
a) The spherically symmetric, asymptotically flat solutions to Eqs~\eqref{f1} to \eqref{f4} do not violate the NEC. b) At the points where $V\geq 0$, the WEC is satisfied and the mass density is positive. c) If the function $r(\rho)-\rho$ admits an expansion in powers of $1/\rho$, then at spatial infinity the spherically symmetric, asymptotically flat solutions to Eqs~\eqref{f1} to \eqref{f4}, whether $V>0$ or $V<0$, do not violate the WEC and the mass density and radial pressure are positive.
\end{corollary}

The consideration of the WEC does not allow for further selection criteria at least at spatial infinity; this is not the case near the center where it is generally possible to select within the same family (P-L, NS or BH solution) those solutions which do not violate the WEC, as will be established shortly below. For that purpose, we focus on the case of a BH solution then tabulate the results for the other solutions.

We assume the conditions~\eqref{s2} \& \eqref{s2b} hold and use Eqs~\eqref{a1} \& \eqref{v11}. For a BH solution $c-3m<0$, $\epsilon<0$ \& $s_1^2>1+s_2^2+s_3^2$. If $r''(c)\neq 0$ ($r''(c)<0$), in the limit $x\rightarrow 0^+$ Eq~\eqref{ec5} leads to $T^{\mu}{}_{\nu}u_{\mu}u^{\nu}\propto (c-3m)r''(c)/x^2$, so the WEC is satisfied. If $r''(c)=0$, $T^{\mu}{}_{\nu}u_{\mu}u^{\nu}\propto (c-3m)r'''(c)/x$ so that the WEC is satisfied if $r'''(c)<0$. Now, if $r''(c)=0$ \& $r'''(c)=0$, we have $T^{\mu}{}_{\nu}u_{\mu}u^{\nu}\propto -(c-3m)r^{(4)}(c)\ln x$ and the WEC is satisfied if $r^{(4)}(c)<0$. Finally, if $r''(c)=0$, $r'''(c)=0$ \& $r^{(4)}(c)=0$, $T^{\mu}{}_{\nu}u_{\mu}u^{\nu}\propto -L$ and the WEC is satisfied if $L<0$.

Table~\ref{Tab2} shows the conditions for which the WEC is satisfied near the center. There is no case of validity of the WEC for a NS with $c-3m>0$ \& $r''(c)=0$ (case 1b) in Table~\ref{Tab2}): the WEC is violated $\forall\;r'''(c)\neq 0$. In fact, under these conditions we have $T^{\mu}{}_{\nu}u_{\mu}u^{\nu}\propto (c-3m)(1-3s_1^2-s_2^2-s_3^2)r'''(c)/x$. The WEC is violated if $1-3s_1^2-s_2^2-s_3^2$ \& $r'''(c)$ have opposite signs. Since for a NS $\epsilon =1$, $0\leq s_1^2<1$ \& $0<s_1^2+s_2^2+s_3^2<1$, this implies $-2s_1^2<1-3s_1^2-s_2^2-s_3^2<1-2s_1^2\leq 1$, thus $1-3s_1^2-s_2^2-s_3^2$ may have both signs. For instance, for a four-velocity vector or an observer with $s_1=1/2$ \& $s_2=s_3=0$, $1-3s_1^2-s_2^2-s_3^2=1/4$ and the WEC is violated if $r'''(c)<0$; for an other observer with $s_1=1/\sqrt{2}$ \& $s_2=s_3=0$, $1-3s_1^2-s_2^2-s_3^2=-1/2$ and the WEC is violated if $r'''(c)>0$. Similarly, there is no case of validity of the WEC for a NS with $c-3m=0$ \& $r''(c)<0$ (case 2a) in Table~\ref{Tab2}).
\begin{table}
\caption{{\small Static, spherically symmetric, asymptotically flat solutions to Eqs~\eqref{f1} to~\eqref{f4}. Validity of the WEC near the center for P-L, NS \& BH solutions. $r(\rho)$ is subject to conditions~\eqref{s2} \& \eqref{s2b}.}}
\label{Tab2}
{\small
\begin{tabular}{|l|l|l|}
\noalign{\smallskip}
  \hline
  \textsc{Particle-Like} & \textsc{Naked Singularity:} & \textsc{Black Hole:} \\
   \textsc{Solution:} $c-3m=0$, & $c-3m\geq 0$ & $c-3m<0$ \\
  $r''(c)=0$, $r'''(c)=0$   &  &  \\
  \hline
   & 1) $c-3m>0$: & \\
   & a) $r''(c)<0$ & a) $r''(c)<0$ \\
   &  & b) $r''(c)=0$ \& $r'''(c)<0$ \\
   & c) $r''(c)=0$, $r'''(c)=0$ & c) $r''(c)=0$, $r'''(c)=0$ \\
   & \& $r^{(4)}(c)>0$ & \& $r^{(4)}(c)<0$ \\
 d) $L<0$  & d) $r''(c)=0$, $r'''(c)=0$, & d) $r''(c)=0$, $r'''(c)=0$, \\
   & $r^{(4)}(c)=0$ \& $L<0$ & $r^{(4)}(c)=0$ \& $L<0$ \\
  \hline
  & 2) $c-3m=0$: & \\
  & b) $r''(c)=0$ \& $r'''(c)>0$ & \\
  \hline
\end{tabular}}
\end{table}

\section{Further exact solutions}\label{Sec4}
The purpose of this section is not to discuss in detail the physical properties of the derived solutions; rather it is to show how to construct predefined solutions using the first five corollaries. The existence of a critical mass $m_c$ and the validity the WEC at spatial infinity and that of the NEC will be the cases since for all solutions derived in this section and those discussed earlier, Eqs~\eqref{R1} \& \eqref{R2} to~\eqref{R5}, the function $r(\rho)-\rho$ admits an expansion in powers of $1/\rho$ at spatial infinity. We will use the inverse problem method by which we select a function $r(\rho)$ and integrate Eqs~\eqref{A5} \& \eqref{P2} to obtain $A(\rho)$ \& $\phi(\rho)$, respectively, and derive $V(\rho)$ from  Eq~\eqref{V4}. For the three solutions presented below, the expressions for $A(\rho),\;V(\rho)\;\&\;\phi(\rho)$, if they are available, are given in the corresponding part of Appendix B.

\subsection{Black holes \& naked singularities with $V$ negative at spatial infinity}\label{suse1}
Let $r(\rho)$ be defined by
\begin{equation}\label{s1r}
    r(\rho) = \rho - \frac{c^2}{\rho}\,,\;c>0\,.
\end{equation}
Eqs~\eqref{A5}, \eqref{V4} \& \eqref{P2} lead to Eqs~\eqref{su1} for $A,\;\phi\;\&\;V$. By corollaries 4 \& 5, the solution is either a BH or a NS. Using Eq~\eqref{V1} with $a_1=-c^2$ \& $a_2=0$ we conclude that $V$ is always negative at spatial infinity for positive masses ($m>m_c=0$) whatever the solution is. At the center $c$, $A$ \& $V$ have opposite signs: $A\simeq (c-3m)/(6\rho)$ \& $V\simeq -(c-3m)/(12c\rho^2)$.

\begin{figure}[h]
\centering
  \includegraphics[width=\textwidth]{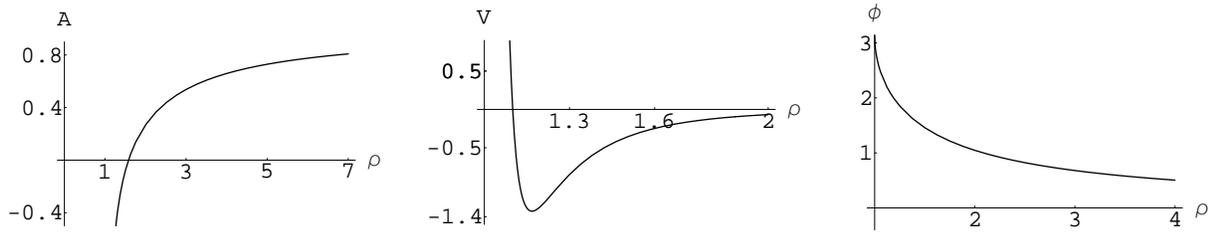}\\
  \caption{\footnotesize{A BH solution with center $c=1$ \& mass $m=2/3$ derived choosing $r(\rho)$ of the form~\eqref{s1r}. The horizon is at $\rho_{\text{h}} =1.5806$ \& the minimum of $V$ is at $\rho_{\text{min}} = 1.16804$. $\phi(c)=\pi$.}}\label{Fig2}
\end{figure}

\begin{figure}[ht]
\centering
 \includegraphics[width=\textwidth]{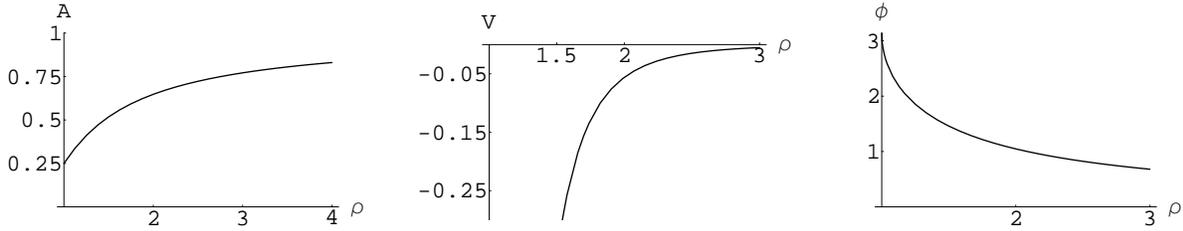}\\
  \caption{\footnotesize{A NS solution with center $c=1$ \& mass $m=1/3$ derived choosing $r(\rho)$ of the form~\eqref{s1r}. $\phi(c)=\pi$ \& $\lim_{\rho\to c^+}A(\rho)=1/4$.}}\label{Fig3}
\end{figure}

Fig.~\ref{Fig2} depicts a BH solution for $c=1\;\&\;m=2/3$. The horizon is at $\rho_{\text{h}} =1.5806$ ($c<\rho_{\text{h}}<3m$ by corollary 2), the potential $V$ crosses the axis at $\rho_1 = 1.09995$ and reaches its minimum value at $\rho_{\text{min}} = 1.16804$. The behavior of the solution is as stated in corollary 4.
Fig.~\ref{Fig3} depicts a NS solution for $c=1\;\&\;m=1/3$ the potential $V$ of which is always negative. The behavior of the solution is as stated in corollary 5.

\subsection{Black holes \& naked singularities with $V$ having both signs at spatial infinity}\label{suse2}
In the previous example~\eqref{s1r} it was not possible to derive a BH solution with $V$ positive at spatial infinity or a NS with $V$ positive near the center. We realize that by making another choice for $r(\rho)$
\begin{equation}\label{s2r}
    r(\rho) = \rho - \frac{b^2}{4(\rho-b)}\,,\;b>0\,,\;\rho \geq c=\frac{\sqrt{2}+1}{2}\,b>0\,,
\end{equation}
where $b$ is any positive number and the center $c=(\sqrt{2}+1)b/2$. At spatial infinity we have $r(\rho)= \rho - (b^2/4)/\rho - (b^3/4)/\rho^2+O(1/\rho^3)$. By corollary 6, the potential is positive as $\rho\to\infty$ if $m<m_c=3/4$, regardless of the nature of the solution. A BH with $c/3<m<m_c$ or a NS with $m<\min(c/3,m_c)$ has its potential positive as $\rho\to\infty$. At the center $c$, on can show using Eqs~\eqref{a1} \& \eqref{v11}, with $r(\rho)$ given by~\eqref{s2r}, that $A$ \& $V$ have always opposite signs for all $m>0$ \& $c>0$.

Eqs~\eqref{A5}, \eqref{V4} \& \eqref{P2} lead to Eqs~\eqref{su2} to~\eqref{su22} for $A,\;V\;\&\;\phi$. By corollaries 4 \& 5, the solution is either a BH or a NS.

\begin{figure}[ht]
\centering
  \includegraphics[width=\textwidth]{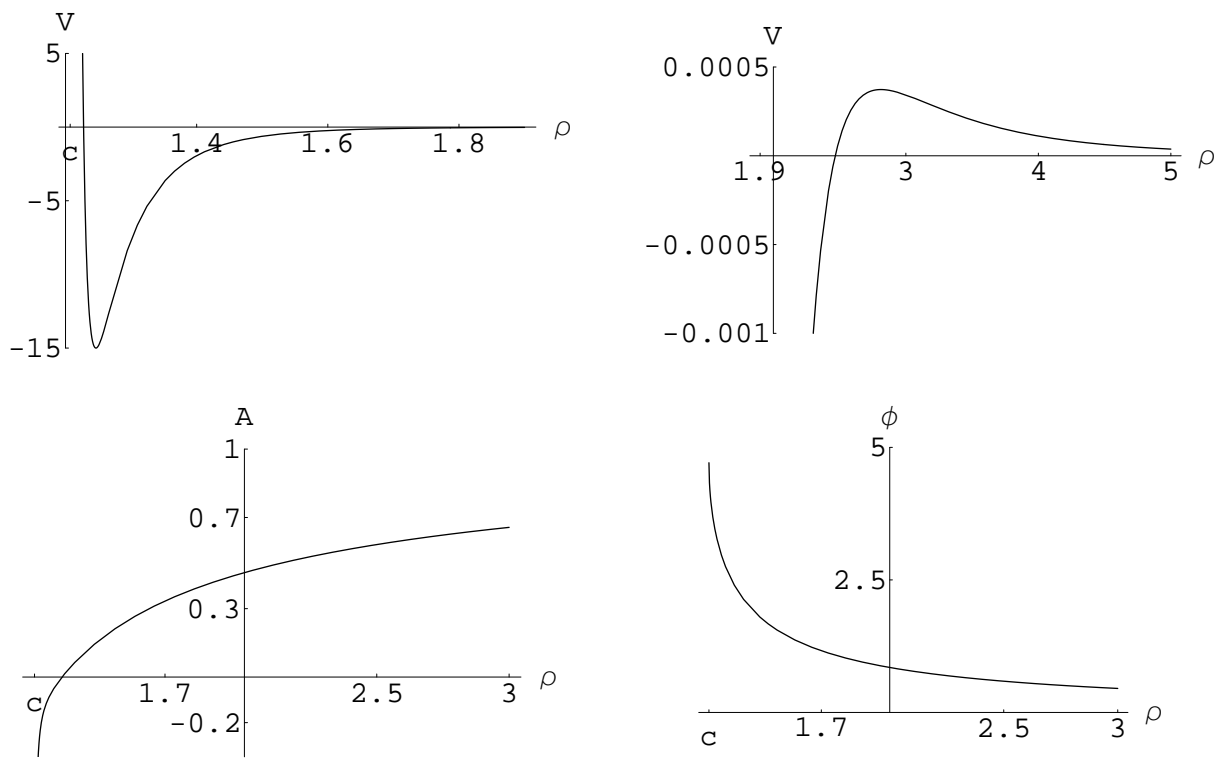}\\
  \caption{\footnotesize{A BH solution with center $c=(\sqrt{2}+1)/2$ \& mass $m=1/2$ derived choosing $r(\rho)$ of the form~\eqref{s2r}. The horizon is at $\rho_{\text{h}} =1.31091$ and the minimum and maximum values of $V$ are at $\rho_{\text{min}} =1.2464$ \& $\rho_{\text{max}} =2.81174$, respectively. $\phi(c)=3\pi/2$. Two plots of $V$ are shown for $c\leq \rho \leq 1.9$ \& $1.9\leq \rho \leq 5$.}}\label{Fig4}
\end{figure}

\begin{figure}[ht]
\centering
  \includegraphics[width=\textwidth]{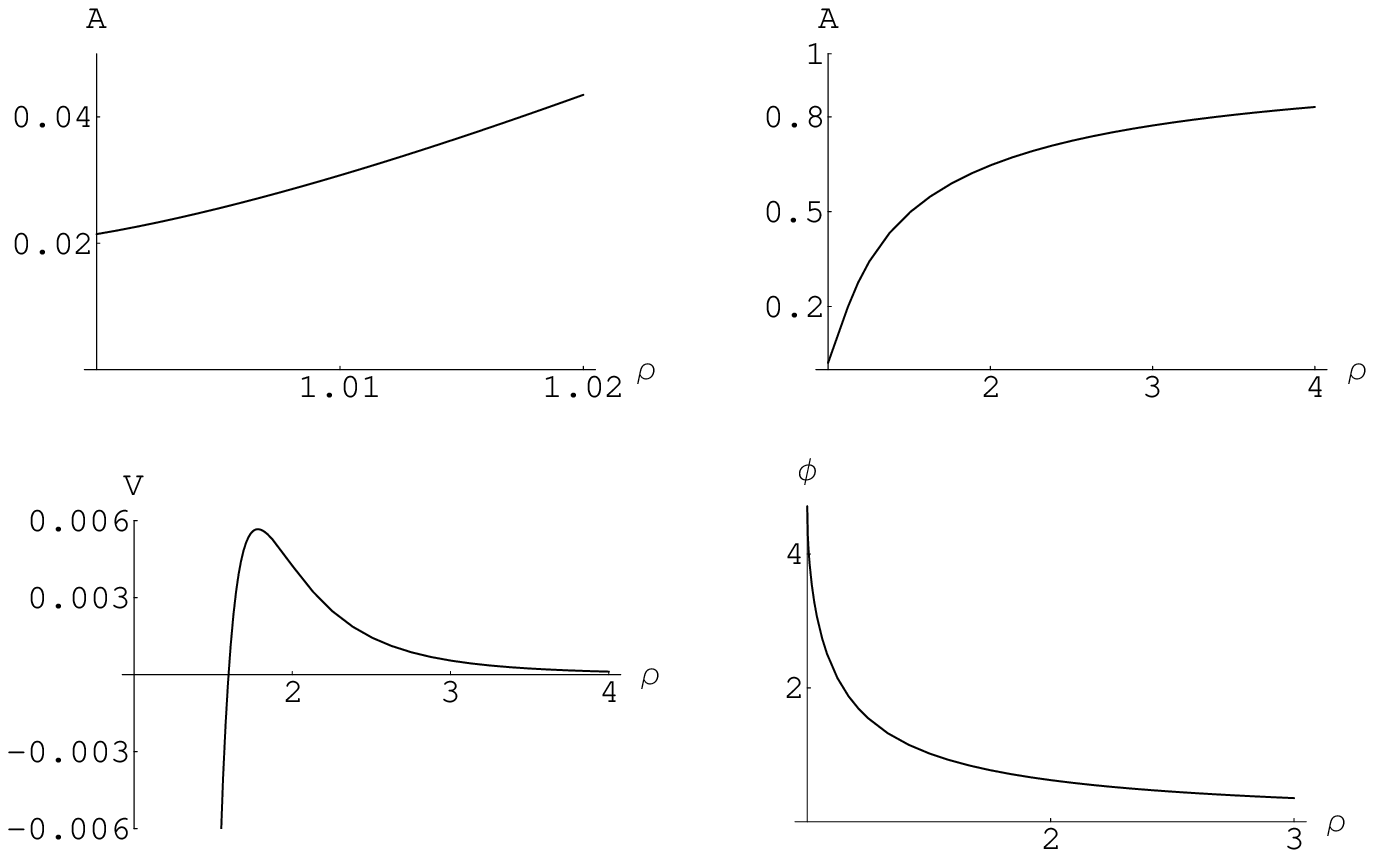}\\
  \caption{\footnotesize{An extreme NS solution with center $c=1$ \& mass $m=1/3$ derived choosing $r(\rho)$ of the form~\eqref{s2r}. $\phi(c)=3\pi/2$. Two plots of $A$ are shown and $\lim_{\rho\to c^+}A(\rho)=(3-\sqrt{8})/8=0.0214466$.}}\label{Fig5}
\end{figure}

Taking $m=1/2$ \& $c=(\sqrt{2}+1)/2$ ($b=1$), one obtains the BH solution depicted in Fig.~\ref{Fig4} where the horizon is at $\rho_{\text{h}} =1.31091$ ($c<\rho_{\text{h}}<3m$ by corollary 2). The potential crosses the axis at $\rho_1 =1.22754$ \& $\rho_2 =2.46669$ and reaches its minimum and maximum values at $\rho_{\text{min}} =1.2464$ \& $\rho_{\text{max}} =2.81174$, respectively. The solution behaves as stated in corollary 4.
Taking $m=1/3$ \& $c=1$, one obtains the extreme NS depicted in Fig.~\ref{Fig5}. The function $A$ is first concave up then concave down, which is obvious from the two plots of $A$ in Fig.~\ref{Fig5}. The solution behaves as stated in corollary 5.

\subsection{A one-parameter particle-like solution}\label{suse3}
In the two previous example~\eqref{s1r} \& \eqref{s2r} it was not possible to derive a P-L solution. We realize that by making a third choice for $r(\rho)$. We choose $r(\rho)$ of the form~\eqref{R1} with $n=4$ (the simplest case). In order to fulfill the requirements of the theorem we take $c=3m$. Since the Einstein-scalar theory is scale invariant, we restrict ourselves to the case $c=3m=1$. Once the expression of $A(\rho)$ is obtained for $c=3m=1$, one generates the general expression of $A$ for $c=3m\neq 1$ by replacing $\rho$ by $\rho/c$ and $m$ by $m/c$ in the r.h.s of $A(\rho)$. Hence, we start with
\begin{equation}\label{s3r}
    r(\rho) = \frac{4}{3}\,(\rho - 1) - \frac{1}{3}\,\frac{(\rho - 1)^4}{\rho^{3}}\,,\;\rho \geq 1\,.
\end{equation}
Eq~\eqref{A5} leads to Eq~\eqref{su3} for $A$. The solution behaves as stated in the theorem with $\lim_{\rho\to c^+}V(\rho)=-3\lim_{\rho\to c^+}$ $A''(\rho)/2=-1.81735$ (Eq~\eqref{le}). By corollary 6, positive masses ($m>m_c=-1/2$) have negative potential at spatial infinity.
\begin{figure}[ht]
\centering
 \includegraphics[width=\textwidth]{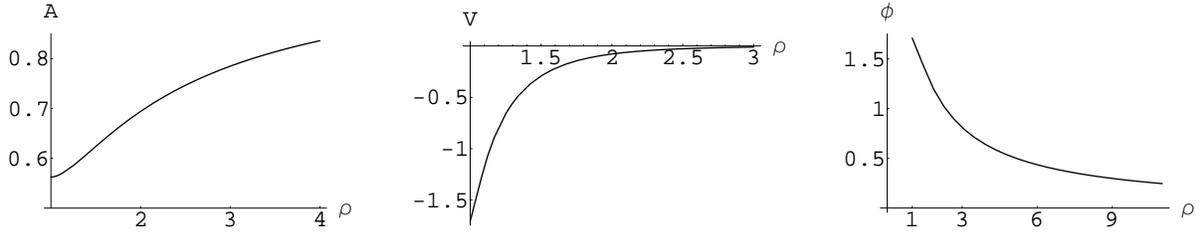}\\
  \caption{\footnotesize{A P-L solution with center $c=1$ \& mass $m=1/3$ derived choosing $r(\rho)$ of the form~\eqref{R1} with $n=4$ (Eq~\eqref{s3r}). $\phi(c)=1.71007$, $\lim_{\rho\to c^+}A(\rho)=9/16=0.5625$ \& $\lim_{\rho\to c^+}V(\rho)=-3\lim_{\rho\to c^+}A''(\rho)/2=-3\times 1.211569/2=-1.81735$.}}\label{Fig6}
\end{figure}
\begin{figure}[ht]
\centering
 \includegraphics[width=\textwidth]{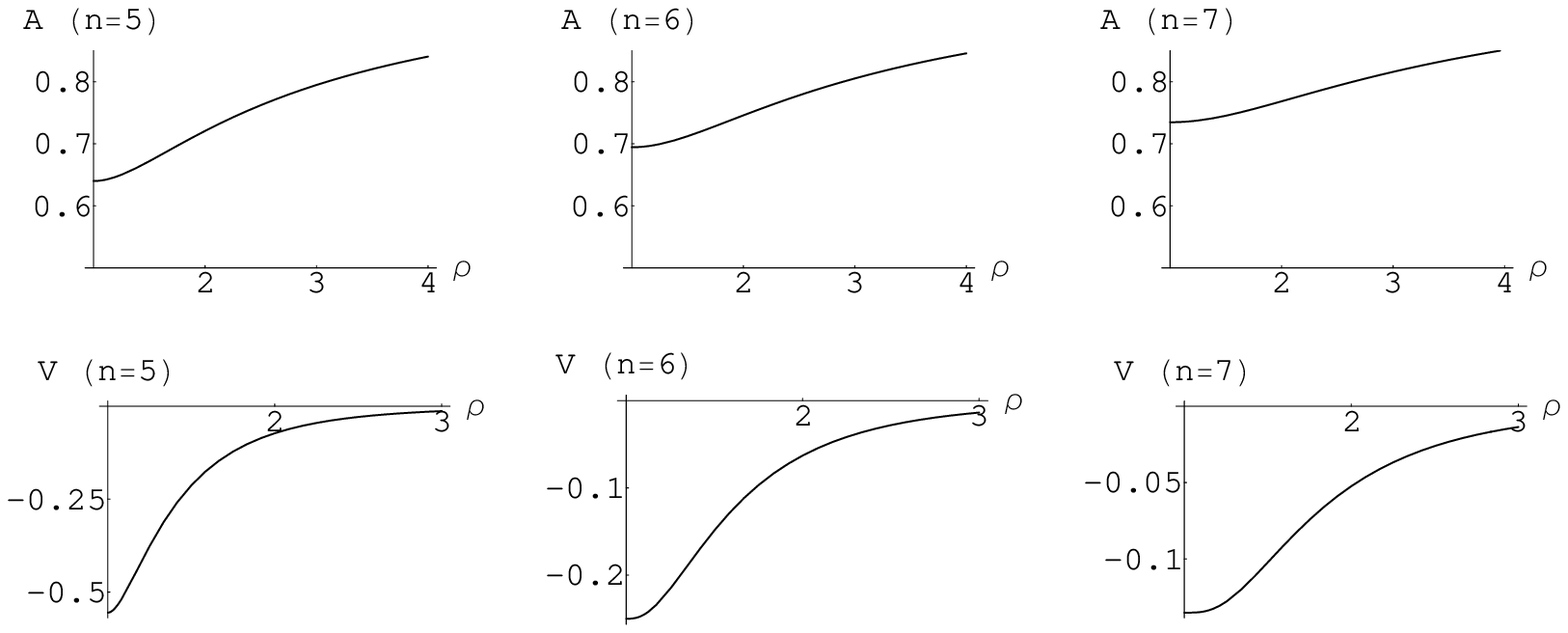}\\
  \caption{\footnotesize{P-L solutions with center $c=1$ \& mass $m=1/3$ derived choosing $r(\rho)$ of the form~\eqref{R1} with $n=5,\,6\,\&\,7$ (left to right).}}\label{Fig7}
\end{figure}

The P-L solutions derived from~\eqref{R1} for $n\geq 4$ have similar behaviors with $V$ negative everywhere. Plots of $A$ \& $V$ for $n=5,\,6\,\&\,7$ and the same mass and center are shown in Fig.~\ref{Fig7}. As $n\to\infty$, $A$ \& $V$ approach their Minkowskian limits. Such a behavior is desirable in theories where the effects of the scalar field are confined.

\section{Conclusion}
The selection criteria we have established in the first five corollaries and theorem have been used to help discussing and constructing explicit solutions of particular interest. In section~\ref{Sec2} we provided a unified formula for $r(\rho)$ (Eq~\eqref{R1}) which generates different P-L, NS \& BH solutions (behavior of $V$ and horizon location (for BHs) depend on $n$). In section~\ref{Sec3} we have shown the existence of new kind of BHs (Eq~\eqref{R2}) where the metric $A$ is singular at the center only and the matter fields $V$ \& $\phi$ are regular everywhere. A parallel conclusion applies to the existence of NSs (Eq~\eqref{R3}).

We have shown that the NEC (any $V$) and the WEC (only $V\geq 0$) are always satisfied. Particularly, if $r-\rho$ admits an expansion in terms of $1/\rho$, then 1) there exists a sort of critical mass (if $a_1\neq 0$) such that $V$ at spatial infinity \& $(m_c-m)$ have the same sign, regardless the nature of the solution. 2) The solutions to the field equations obey the WEC at spatial infinity (any $V$). If the derivatives of $r$ at the center and $L$ have specific signs, the WEC is satisfied there.

We have constructed BHs, Fig.~\ref{Fig2} (resp. Fig.~\ref{Fig4}), the potential function of which has one local minimum value (resp. one local minimum and one local maximum values). The two BHs behave differently at spatial infinity. We have also discussed NSs, Fig.~\ref{Fig3} (resp. Fig.~\ref{Fig5}), the potential function of which is negative everywhere (resp. has both signs). We have constructed families (a family for each value of $n\geq 4$ in~\eqref{R1}) of one-parameter P-L solutions,  Figures~\ref{Fig6} \& \ref{Fig7}, which have negative potential everywhere. Each family of P-L solution is different from that constructed in Ref~\cite{N08}.

\section*{Appendix A}
\appendix
\def\theequation{A.\arabic{equation}}
\setcounter{equation}{0}
Appendix A is divided into two parts and its purpose is three-fold. In Part 1, where $c=3m$ and $r(\rho)$ satisfies the conditions~\eqref{subj1} \& \eqref{subj3}, we prove that a) $\lim_{\rho\to c^+}A'(\rho)$ is finite and b) if $15r''^{\,2}(c)-4r'(c)r'''(c)=0$ (particularly if $r''(c)=0$ \& $r'''(c)=0$) then $\lim_{\rho\to c^+}A''(\rho)$ is finite and provide a value for the latter. In Part 2, where $c\neq 3m$ and $r(\rho)$ satisfies the conditions~\eqref{s2} \& \eqref{s2b}, we derive formulas~\eqref{a1}, \eqref{v11} \& \eqref{p11} for $A,\;V\;\&\;\phi$.

\subsection*{1) \textsc{Part 1:} $c=3m$}
We introduce the variable $x=\rho-c$. While $r(x+c)$ is supposed to be four times differentiable by hypotheses~\eqref{subj1} \& \eqref{subj3}, $r^{(4)}(x+c)$ need be defined on a small half-open domain including the center $x=0$. By Taylor's Theorem there exists a number $0<\xi_1(x)<x$ such that
\begin{equation}\label{ap1}
    r(x+c)=r'(c)x+\frac{r''(c)}{2}\,x^2+\frac{r'''(\xi_1(x))}{6}\,x^3\,.
\end{equation}
Now, we introduce the continuous function
\begin{equation*}
    \tau(x)=\left\{\begin{array}{ll}
    r(x+c)/[r'(c)x]\,, & \text{if}\; x>0 \\
    1\,, & \text{if}\; x=0
                     \end{array}\,.\right .
\end{equation*}
Since $r(x+c)$ has only one zero: $r(c)=0$, the function $\tau(x)$ has no zero in the half-open domain $x\in [0,\infty)$. The derivatives of $\tau$ up to order $3$ exist and are finite for $x\in [0,\infty)$. Particularly, the first derivative of $\tau$ depends on $r''$ and its second derivative depends on $r'''$: $\tau'(0)=r''(c)/2r'(c)\;\; \& \;\;\tau''(0)=r'''(c)/3r'(c)$.
For all $x\geq 0$ we have
\begin{equation}\label{app3}
    r(x+c)=r'(c)x\tau(x)\,.
\end{equation}

Next, consider the continuous function $\sigma(x)$ defined by
\begin{equation*}
    \sigma(x)=\left\{\begin{array}{ll}
   x^4/r^4(x+c)=1/[r'^{\,4}(c)\tau^4(x)]\,, & \text{if}\; x>0 \\
    1/r'^{\,4}(c)\,, & \text{if}\; x=0
                     \end{array}\,.\right .
\end{equation*}
The function $\sigma$ and its first three derivatives are finite and continuous on the domain $x\in [0,\infty)$ and have no poles in it. By Taylor's Theorem there exists a number $0<\xi_2(x)<x$ such that
\begin{equation}\label{ap4}
\sigma(x)=\frac{1}{r'^{\,4}(c)}-\frac{2r''(c)}{r'^{\,5}(c)}\,x
    +\frac{15r''^{\,2}(c)-4r'(c)r'''(c)}{6r'^{\,6}(c)}\,x^2+\frac{\sigma'''(\xi_2(x))}{6}\,x^3\,.
\end{equation}
Notice how the derivative of order $n-1$ of $\sigma$ depends on the derivative of order $n$ of $r$.

We introduce an arbitrary constant $x_0$: $0<x<x_0<\infty$ and let $C_1$, $C_3$, $C_2$, $\delta$ \& $\lambda$ be the constants
\begin{align*}
& C_1 = -2\int_{\infty}^{x_0}\frac{y}{r^4(y+c)}
    \,\text{d}y = -2\int_{\infty}^{x_0}\frac{\sigma(y)}{y^3}
    \,\text{d}y\,,\quad C_3 = \int_{\infty}^{x_0}\frac{1}{r^4(y+c)}
    \,\text{d}y = \int_{\infty}^{x_0}\frac{\sigma(y)}{y^4}
    \,\text{d}y\,,\\
& C_2=\frac{1}{2r'^{\,4}(c)x_0^2}-\frac{2r''(c)}{r'^{\,5}(c)x_0}
    -\frac{15r''^{\,2}(c)-4r'(c)r'''(c)}{6r'^{\,6}(c)}\,\ln x_0 - \frac{S(x_0)}{6}\,,\\
& \delta = \frac{15r''^{\,2}(c)-4r'(c)r'''(c)}{6r'^{\,6}(c)}\,,
    \quad \lambda = \frac{10r'(c)r''(c)r^{(3)}(c)-15 r''^{\,3}(c)-r'^{\,2}(c)r^{(4)}(c)}{6r'^{\,7}}\,.
\end{align*}
With $c=3m$ we rewrite~\eqref{A5} using the new variable $x$ and the constant $C_1$
\begin{equation}\label{ap7}
    A = C_1r^2-2\left(\int_{x_0}^{x}\frac{y}{r^4(y+c)}
    \,\text{d}y\right)r^2\,,\;\;(0<x<x_0<\infty)\,.
\end{equation}
For $x>0$ ($y>0$), $y/r^4(y+c)=\sigma(y)/y^3$. Thus using~\eqref{app3} \& \eqref{ap4} in~\eqref{ap7} leads to
\begin{eqnarray}
  A &=& \left(C_1-2C_2-\frac{S(x)}{3}\right)r'^{\,2}(c)x^2\tau^2(x) + \frac{\tau^2(x)}{r'^{\,2}(c)}
  \nonumber  \\
\label{ap8}& &  -\frac{4r''(c)}{r'^{\,3}(c)}\,x\tau^2(x) - \frac{15r''^{\,2}(c)-4r'(c)r'''(c)}{6r'^{\,6}(c)}\,\tau^2(x)x^2\ln x\,,\\
\text{with}& & S(x)=S(x_0)+\int_{x_0}^x\sigma'''(\xi_2(y))\,\text{d}y\,.\nonumber
\end{eqnarray}
Since $\sigma'''$ has no pole in the domain $x\in [0,\infty)$, $S(x)$ is finite and continuous on it.

Taking the derivative of~\eqref{ap8} with respect to $x$ then the limit as $x\to 0^+$ we obtain Eq~\eqref{A10}. The non-vanishing contributions come from the second term in the r.h.s of~\eqref{ap8}, which has a derivative $2\tau\tau'/r'^{\,2}(c)\to r''(c)/r'^{\,3}(c)$ as $x\to 0^+$, and the third one the derivative of which approaches $-4r''(c)/r'^{\,3}(c)$ as $x\to 0^+$. We have thus proven that $\lim_{\rho\to c^+}A'(\rho)=A'(c)=-3r''(c)/r'^{\,3}(c)$ is finite whenever $c=3m$.

It is obvious from~\eqref{ap8} that a necessary condition for $\lim_{\rho\to c^+}A''(\rho)$ to be finite is
\begin{equation}\label{ap11}
    15r''^{\,2}(c)-4r'(c)r'''(c)=0\,.
\end{equation}
We assume that~\eqref{ap11} holds. Now, $A''(x+c)$ is provided by~\eqref{f4}
\begin{align}
\label{ap12} & A''=\frac{(r^2)''A-2}{r^2}\,,\\
\label{ap13} \;\text{with (use~\eqref{app3}):}\;\;& (r^2)''=2r'^{\,2}(c)[\tau^2+4x\tau'\tau+x^2\tau'^{\,2}+x^2\tau\tau'']\,.
\end{align}
All functions in the r.h.s of~\eqref{ap13} are continuous and finite for $x\in [0,\infty)$ ($\tau''$ depends on $r'''$ which exists and is finite by hypothesis). Using~\eqref{app3}, \eqref{ap8} \& \eqref{ap13} in~\eqref{ap12}, one sees that the only non-finite contributions to $\lim_{\rho\to c^+}A''(\rho)$ may come from the second and third terms in~\eqref{ap8}. These two contributions to $A''$ are
\begin{equation*}
\left(2(\tau^4-1)+8x\tau'\tau^3-\frac{8r''(c)}{r'(c)}\,x\tau^4\right)/r^2 \to \frac{4r''(c)}{r'^{\,3}(c)x}+\frac{4r''(c)}{r'^{\,3}(c)x}-\frac{8r''(c)}{r'^{\,3}(c)x}=0\,.
\end{equation*}
Thus, we have proven that $\lim_{\rho\to c^+}A''(\rho)$ is finite whenever condition~\eqref{ap11} is satisfied.

As we have seen earlier, the potential $V$ is finite at the center only if $r''(c)=0$ \& $r'''(c)=0$ (Eqs~\eqref{V5} \& \eqref{le}). In this case Eq~\eqref{ap11} is satisfied. We postpone the derivation of a formula for $A''(c)$ to Part 2. We will see that for $c=3m$, $r''(c)=0$ \& $r'''(c)=0$ , $A''(c)$ reduces to $L$
\begin{equation}\label{ap15}
    L=2\lim_{x_0\to 0^+}\bigg(-2r'^{\,2}(c)\int_{\infty}^{x_0}\frac{y\,\text{d}y}{r^4(y+c)}-\frac{1}{r'^{\,2}(c)x_0^2}
    \bigg)\,,\;\text{which is just Eq}\; \eqref{new}\,.
\end{equation}

\subsection*{2) \textsc{Part 2:} $c\neq 3m$}
$r(x+c)$ is supposed to be five times differentiable by hypotheses~\eqref{s2} \& \eqref{s2b}. As before $r^{(5)}(x+c)$ need be defined on a small half-open domain including the center $x=0$. Replace~\eqref{ap1} \& \eqref{ap4}, respectively, by
\begin{align}
\label{nw1}r(x+c)=& r'(c)x+\frac{r''(c)}{2}\,x^2+\frac{r'''(c)}{6}\,x^3
    +\frac{r^{(4)}(c)}{24}\,x^4+\frac{r^{(5)}(\xi_1(x))}{120}\,x^5\,,\\
\label{nw2}\sigma(x)=& \frac{1}{r'^{\,4}(c)}-\frac{2r''(c)}{r'^{\,5}(c)}\,x
    +\delta\,x^2 +\lambda\,x^3 +\frac{\sigma^{(4)}(\xi_2(x))}{24}\,x^4\,,
\end{align}
($\sigma^{(4)}(x=0)$ depends on $r^{(5)}(c)$). $A$ (Eq~\eqref{A5}) takes the form
\begin{align}
\label{nw4b}A=& -2r^2(x+c)\left(\int_{\infty}^{x}\frac{y\,\text{d}y}{r^4(y+c)}\right)-
2(c-3m)r^2(x+c)\left(\int_{\infty}^{x}\frac{\text{d}y}{r^4(y+c)}\right)\\
\label{nw4}= & [C_1-2(c-3m)C_3]r^2(x+c) - 2r^2(x+c)\int_{x_0}^{x}\frac{\sigma(y)\,\text{d}y}{y^3} - 2(c-3m)r^2(x+c)\int_{x_0}^{x}\frac{\sigma(y)\,\text{d}y}{y^4}\,.
\end{align}
Define the new continuous functions $I_1(x)$ \& $I_2(x)$ by
\begin{equation}\label{nw5}
    I_1(x) = I_1(x_0) + \frac{1}{24}\int_{x_0}^{x}y\sigma^{(4)}(\xi_2(y))
    \,\text{d}y\,, \quad I_2(x) = I_2(x_0) + \frac{1}{24}\int_{x_0}^{x}\sigma^{(4)}(\xi_2(y))
    \,\text{d}y\,.
\end{equation}
Replacing $\sigma(y)$ in~\eqref{nw4} by the r.h.s of~\eqref{nw2} and using~\eqref{nw5}, we arrive at
\begin{align}
    A = & [C_1-2(c-3m)C_3]r^2(x+c) - 2r^2(x+c)\bigg[I_1(x)-\frac{1}{2r'^{\,4}(c)x^2}
    +\frac{2r''(c)}{r'^{\,5}(c)x}+\delta\ln x+\lambda x - (x_0\rightarrow x)\bigg]\nonumber\\
\label{nw6}& -2(c-3m)r^2(x+c)\bigg[I_2(x)-\frac{1}{3r'^{\,4}(c)x^3}
    +\frac{r''(c)}{r'^{\,5}(c)x^2}-\frac{\delta}{x}+\lambda\ln x - (x_0\rightarrow x)\bigg]\,,
\end{align}
where ($x_0\rightarrow x$) substitutes $x_0$ for $x$ in the terms within the square parentheses.

In order to obtain the behavior of $A$ \& $V$ near the center (for small $x$), we replace $r(x+c)$, $I_1(x)$ \& $I_2(x)$ in~\eqref{nw6} by the r.h.s of~\eqref{nw1}, $I_1(0)$ \& $I_2(0)$, respectively, then extract the coefficients of $1/x$, $x$, $x^2\ln x$, $x^2$ and the independent term, we arrive first at~\eqref{a1}, then using~\eqref{f2}, we obtain~\eqref{v11}. The derivation of~\eqref{p11} is straightforward. The coefficient of $x^2$ in~\eqref{nw6} reads
\begin{align}
   \frac{A''(c)}{2} = & -2r'^{\,2}(c)\int_{\infty}^{x_0}\frac{\sigma(y)\,\text{d}y}{y^3} -\frac{1}{r'^{\,2}(c)x_0^2}+\frac{4r''(c)}{r'^{\,3}(c)x_0}+
   2r'^{\,2}(c)\delta \ln x_0\nonumber\\
\label{nw7}& +2r'^{\,2}(c)\lambda x_0+2r'^{\,2}(c)[I_1(x_0)-I_1(0)]+\frac{r'''(c)}{3r'^{\,3}(c)}
-\frac{15r''^{\,2}(c)}{4r'^{\,4}(c)}\\
& +(c-3m)\bigg[-2r'^{\,2}(c)\int_{\infty}^{x_0}\frac{\sigma(y)\,\text{d}y}{y^4}
-\frac{2}{3r'^{\,2}(c)x_0^3}+\frac{2r''(c)}{r'^{\,3}(c)x_0^2}-\frac{2r'^{\,2}(c)\delta}{x_0}\nonumber\\
& 2r'^{\,2}(c)\lambda \ln x_0+2r'^{\,2}(c)[I_2(x_0)-I_2(0)]+2\delta r'(c)r''(c)+\frac{r^{(4)}(c)}{18r'^{\,3}(c)}
-\frac{r''^{\,3}(c)}{2r'^{\,5}(c)}-\frac{5r''(c)r'''(c)}{9r'^{\,4}(c)}\bigg]\,.\nonumber
\end{align}
$A''(c)$ depends apparently on $x_0$, however, using Eqs~\eqref{nw2} \& \eqref{nw5} it is straightforward to obtain $\text{d}[A''(c)]/\text{d}x_0$ $=0$.
Hence, $A''(c)$ is independent of $x_0$ and is finite since every function or constant in the r.h.s of~\eqref{nw7} is finite. In the limit $x_0\to 0^+$, many terms drop from the r.h.s of~\eqref{nw7} and the remaining terms provide the desired expression for $A''(c)$
\begin{align}
\label{ap16}\frac{A''(c)}{2} = & \lim_{x_0\to 0^+}\bigg[-2r'^{\,2}(c)\int_{\infty}^{x_0}\frac{\sigma(y)\,\text{d}y}{y^3} -\frac{1}{r'^{\,2}(c)x_0^2}+\frac{4r''(c)}{r'^{\,3}(c)x_0}+
   2r'^{\,2}(c)\delta \ln x_0\bigg]\\
& +(c-3m)\lim_{x_0\to 0^+}\bigg[-2r'^{\,2}(c)\int_{\infty}^{x_0}\frac{\sigma(y)\,\text{d}y}{y^4}
-\frac{2}{3r'^{\,2}(c)x_0^3}+\frac{2r''(c)}{r'^{\,3}(c)x_0^2}-\frac{2r'^{\,2}(c)\delta}{x_0}+ 2r'^{\,2}(c)\lambda \ln x_0\bigg]\nonumber\\
& +\frac{r'''(c)}{3r'^{\,3}(c)}-\frac{15r''^{\,2}(c)}{4r'^{\,4}(c)} + (c-3m)\bigg[2\delta r'(c)r''(c)+\frac{r^{(4)}(c)}{18r'^{\,3}(c)}
-\frac{r''^{\,3}(c)}{2r'^{\,5}(c)}-\frac{5r''(c)r'''(c)}{9r'^{\,4}(c)}\bigg]\,,\nonumber
\end{align}
where the limit in the r.h.s depends on the particular form of $r(x+c)$ and cannot be obtained in closed form. If $c-3m=0$, $r''(c)=0$ \& $r'''(c)=0$, Eq~\eqref{ap16} reduces to~\eqref{ap15}.

\section*{Appendix B}
\appendix
\def\theequation{B.\arabic{equation}}
\setcounter{equation}{0}
This section is divided into three parts. In Part 1 to Part 3, we show the expressions of $A(\rho),\;V(\rho)\;\&\;\phi(\rho)$, if they are available, corresponding to subsections~\ref{suse1} to~\ref{suse3}, respectively.

\subsection*{1) \textsc{Part 1:}}
The functions $A(\rho),\;V(\rho)\;\&\;\phi(\rho)$ of subsections~\ref{suse1} take the forms
\begin{eqnarray}
 A &=& 1 - \frac{3m\rho}{8c^2} - \frac{11m}{8\rho} - \frac{c^2}{3\rho^2}\left(\frac{c^2-3m\rho}{c^2-\rho^2}\right) + \frac{3m(c^2-\rho^2)^2\ln\left(\frac{\rho + c}{\rho - c}\right)}{16c^3\rho^2}\,,\quad \phi =\pi - 2\arccos(c/\rho)\,,\nonumber\\
\label{su1} V &=& \frac{15m}{8\rho^3} + \frac{9m}{8c^2\rho} + \frac{5c^6-6c^4\rho^2+3mc^2\rho^3}{3\rho^4(c^2-\rho^2)^2} -  \frac{3m(5c^4+4c^2\rho^2+3\rho^4)\ln\left(\frac{\rho + c}{\rho - c}\right)}{16c^3\rho^4}\,.
\end{eqnarray}

\subsection*{2) \textsc{Part 2:}}
The functions $A(\rho),\;V(\rho)\;\&\;\phi(\rho)$ of subsections~\ref{suse2} take the forms
\begin{eqnarray}
 A &=& \frac{(18 m-11 b) \rho }{32 b^2} + \frac{3 b^2+42 b m-2 (b  +18 m) \rho }{6 (4 \rho ^2-4 b \rho-b^2)} + \frac{(11 b-18 m) \rho ^2 \ell(\rho
)}{32 \sqrt{2} b^3}\nonumber\\
& &+ \frac{106 b+36 m+\sqrt{2}(18 m-11 b) \ell(\rho)}{128 b} + \frac{43 b-210 m-3\sqrt{2}(11 b+18 m) \ell(\rho)}{384 (\rho -b )}\nonumber\\
\label{su2}& &+\frac{264 b m-76 b^2+3\sqrt{2}b(11 b-18 m) \ell(\rho)}{3072 (\rho -b)^2}\,,\\
 V &=& \frac{25 (11 b-18 m)}{96 b^2 (\rho -b)}-\frac{2 [b^2+6 b m+2 (b -6 m) \rho ]}{3 (4 \rho ^2-4 b \rho-b^2)^2} + \frac{30 b m-13 b^2-2(11 b -18 m) \rho }{4 \rho ^2-4 b \rho-b^2}\nonumber\\
& &-\frac{3 (11 b-18 m) \ell(\rho)}{32 \sqrt{2} b^3}+\frac{109 b-318 m-3\sqrt{2}(11 b+18 m) \ell(\rho)}{192 b (\rho -b )^2}\\
& &+\frac{89 b-6 m+3\sqrt{2}(11 b-18 m) \ell(\rho)}{384 (\rho -b)^3}-\frac{5 [(264 b m-76 b^2+3b\sqrt{2}(11 b-18 m) \ell(\rho)]}{3072
(\rho -b)^4}\,,\nonumber\\
\label{su22}\phi &=& \frac{3 \pi }{2} - 4 \arctan\left((\sqrt{2}+1) \sqrt{\frac{2 \rho -(\sqrt{2}+1) b}{2 \rho +(\sqrt{2}-1) b}}\right)\,,
\end{eqnarray}
where $\ell(\rho) = \ln\left(\dfrac{(2+\sqrt{2})\rho-b}{(2-\sqrt{2})\rho-b}\right)=
    \ln\left(\dfrac{(\sqrt{2}+1)\rho-(\sqrt{2}-1)c}{(\sqrt{2}+1)(\rho - c)}\right)$.

\subsection*{3) \textsc{Part 3:}}
The function $A(\rho)$ of subsections~\ref{suse3} is derived here. With our choice~\eqref{s3r}, the integrand in~\eqref{A5} is a rational function of the form $-81\rho^{12}/[(\rho -1)^3(1+\alpha\rho)^4(1+\gamma\rho +\beta^2\rho^2)^4]$, $\alpha =2^{2/3}-1$, $\beta =2^{1/3}+1$, $\gamma =-(2^{2/3}+2)$ \& $\gamma^2 -4\beta^2 <0$. With $\eta(\rho)=1+\alpha\rho$, $\omega(\rho)=1+\gamma\rho +\beta^2\rho^2$ \& $\Delta^2=4\beta^2 -\gamma^2$ we obtain
\begin{align}
& A(\rho) = \left\{-\frac{1}{6}\left(\frac{3a}{(\rho-1)^2} + \frac{i}{\beta^2\omega^3}
+ \frac{2e+3\eta(f+2p\eta)}{\alpha\eta^3}\right) + \frac{h}{\alpha}\left[\ln\left(\frac{\eta}{\sqrt{\omega}}\right)-
\ln\left(\frac{\alpha}{\beta}\right)\right]+ \frac{3\pi k\beta^2\gamma}{\Delta^5}\right . \nonumber\\
& -\frac{10\pi\beta^4(2j\beta^2-i\gamma)}{\Delta^7} -
\frac{\pi(2q\alpha+h\gamma)}{2\alpha\Delta} - \frac{2\pi \nu\beta^2}{\Delta^3}
+ \frac{\pi \mu\gamma}{\Delta^3} - \frac{6\pi l\beta^4}{\Delta^5} - \frac{\mu(2+\gamma \rho)}{\Delta^2\omega} +
\frac{10j\beta^4\omega'}{\Delta^6\omega} +
\frac{l\omega'}{2\Delta^2\omega^2}\nonumber\\
& -\frac{k(2+\gamma \rho)}{2\Delta^2\omega^2} +
\frac{5j\beta^2\omega'}{3\Delta^4\omega^2} - \frac{5i\gamma\omega'}{6\Delta^4\omega^2}
 -\frac{5i\beta^2\gamma\omega'}{\Delta^6\omega} +
\frac{3l\beta^2\omega'}{\Delta^4\omega} - \frac{3k\gamma\omega'}{2\Delta^4\omega} +
\frac{\nu\omega'}{\Delta^2\omega} + \frac{j\omega'}{3\Delta^2\omega^3} -
\frac{i\gamma\omega'}{6\beta^2\Delta^2\omega^3} \nonumber\\
\label{su3}& \left . + \left(\frac{20\beta^4(2j\beta^2-i\gamma)}{\Delta^7} +
\frac{6\beta^2(2l\beta^2-k\gamma)}{\Delta^5} + \frac{2(2\nu\beta^2-\mu\gamma)}{\Delta^3} +
\frac{2q\alpha+h\gamma}{\alpha\Delta}\right) \arctan\left(\frac{\omega'}{\Delta}\right)\right\}2r^2(\rho)\,,
\end{align}
where $\omega'$ is the derivative of $\omega$ with respect to $\rho$ and $a$, $e$, $f$, $h$, $i$, $j$, $k$, $l$, $p$, $\mu$, $\nu$ \& $q$ are constants:
\begin{align*}
& a=-\frac{81}{256},\;e=\frac{3+\beta^2}{768},\;f=-\frac{24+\beta^2}{768},\;
  h=-\frac{5}{768}\,(75-58\beta+11\beta^2),\;i=\frac{1}{64}\,(-33+162\beta-65\beta^2),\;  \\
& j=\frac{1}{64}\,(57-82\beta+25\beta^2),\;
k=\frac{5}{32}\,(12-21\beta+7\beta^2),\;l=\frac{1}{32}\,(-62+71\beta-19\beta^2),\;p=\frac{5}{96}\,(3-\beta),\\
& \mu=\frac{1}{128}\,(-123+128\beta-54\beta^2),\;\nu=\frac{1}{128}\,(167-120\beta+26\beta^2),\;
q=\frac{5}{768}\,(-33-50\beta+11\beta^2)\,.
\end{align*}
$V$ is derived from~\eqref{V4} and $\phi$ from~\eqref{P2}. A computer-aided algebra leads to a sizable analytic expression for $\phi$ depending on the incomplete elliptic integral of the first kind and the complete elliptic integral of the third kind. Though the plot of $\phi$ shown in Fig.~\ref{Fig6} has been drawn using the analytic expression of $\phi$, a numerical evaluation of the integral in~\eqref{P2} is as fast as the numerical evaluations of the two non-elementary elliptic functions of which $\phi$ depends.


\end{document}